\documentstyle{mn}
\input{epsf}
\newcommand{\be}{\begin{eqnarray}}
\newcommand{\ee}{\end{eqnarray}}

\title[Characteristics of Ly-$\alpha$ forest ]
{Statistical characteristics  of observed Ly-$\alpha$ 
forest\\ and the shape of initial power spectrum}
\author[Demia\'nski, Doroshkevich \& Turchaninov]
       {M. Demia\'nski$^{1,2}$,  A.G. Doroshkevich$^{3,4}$
        \& V.Turchaninov$^4$\\  
        $1$Institute of Theoretical Physics,
                       University of Warsaw,
                       00-681 Warsaw, Poland\\
        $2$Department of Astronomy, Williams College,
           Williamstown, MA 01267, USA\\
        $3$Theoretical Astrophysics Center,
          Juliane Maries Vej 30,
          DK-2100 Copenhagen \O, Denmark\\
        $4$Keldysh Institute of Applied Math
                        Russian Academy of Sciences,
                        125047 Moscow,  Russia\\
}

\date{Accepted ...,
      Received ...,
        in original form ... .}

\begin{document}
\maketitle

\begin{abstract}
Properties of about 4 500 observed Ly-$\alpha$ absorbers are 
investigated using the model of formation and evolution of DM 
structure elements based on the modified Zel'dovich theory. 
This model is generally consistent with simulations of 
absorbers formation, describes reasonably well the Large Scale 
Structure observed in the galaxy distribution at small redshifts 
and emphasizes the generic similarity of the LSS and absorbers. 

The simple physical model of absorbers asserts that they are 
composed of DM and gaseous matter. It allows us to estimate 
the column density and overdensity of DM and gaseous components 
and the entropy of the gas trapped within the DM potential wells. 
The parameters of DM component are found to be consistent with 
theoretical expectations for the Gaussian initial perturbations 
with the WDM--like power spectrum. The basic physical 
factors responsible for the absorbers evolution are discussed.   

The analysis of redshift distribution of absorbers confirms 
the self consistency of the adopted physical model, Gaussianity 
of the initial perturbations and allows to estimate the shape of 
the initial power spectrum at small scales what in turn restricts 
the mass of the dominant fraction of DM particles to $M_{DM}
\geq~1.5-5$~keV. Our results indicate a possible redshift 
variations of intensity of the UV background by about a factor
of 2 -- 3  at redshifts $z\sim$~2 -- 3.  
\end{abstract}


\section{Introduction}

One of the most perspective methods to study the processes
responsible for the formation and evolution of the structure 
of the Universe is the analysis of properties of absorbers 
observed in spectra of the farthest quasars. The great 
potential of such investigations was discussed already in 
Oort (1981, 1984) just after Sargent et al. (1980) established 
the intergalactic nature of the Ly-$\alpha$ forest. The available 
Keck and VLT high resolution observations of the forest 
provide a reasonable database and allow to apply statistical 
methods for such investigations.

The essential progress achieved recently through numerous high
resolution simulations of absorbers formation and evolution  
confirms that this process is closely connected with the 
initial power spectrum of perturbations. These results allow 
us to consider the properties of absorbers in the context of 
nonlinear theory of gravitational instability (Zel'dovich 1970; 
Shandarin \& Zel'dovich 1988) and to apply the statistical 
description of structure formation and evolution (Demia\'nski \&
Doroshkevich 1999, 2002; hereafter DD99 \& DD02) to the
Ly-$\alpha$  forest. 

This approach is based on the modified Zel'dovich 
approximation and describes the formation and evolution of DM 
structure elements for the CDM and WDM initial power spectra 
without any smoothing or filtering procedures. It outlines
the structure evolution as a random formation and merging 
of Zel'dovich pancakes, their transverse expansion and/or 
compression and transformation into high density clouds 
and filaments. Later on, the hierarchical merging of 
pancakes, filaments and clouds forms rich galaxy walls 
observed at small redshifts. Main stages of this evolution 
are driven by the initial power spectrum. 

At small redshifts the main results of the statistical 
approach are found to be consistent with characteristics  of the
observed and simulated Large Scale Structure  (Demia\'nski et al. 
2000; Doroshkevich, Tucker \& Allam 2002).  Here we use this
approach for the analysis and interpretation  of the absorbers
observed at large redshifts. We show that  the basic observed
characteristics of Ly-$\alpha$ forest  are also successfully
described in the framework of this  theoretical model.

The application of this approach to Ly-$\alpha$ forest was 
already discussed in Demia\'nski, Doroshkevich \& Turchaninov 
(2001 a,b, hereafter Paper I \& Paper II). Here we improve 
this analysis by using a richer and more refined sample of 
$\sim$ 4500 observed absorbers and a more refined model of 
absorbers. Our approach uses more traditional methods of 
investigation of properties of discrete absorbers rather then
associating them with a continuous non-linear  line-of-sight
density field (see, e.g., Weinberg et al. 1998; McDonald et 
al. 2000; Croft et al. 2001; Croft et al. 2002). It allows 
to reach more clarity in the description of formation and 
evolution of absorbers and to reveal its strong link with 
the initial power spectrum. It allows also to separate several 
subpopulations of absorbers and to discuss their evolutionary 
history. 

The composition and spatial distribution of observed absorbers 
is complicated and at low redshifts a significant number of 
stronger Ly-$\alpha$ lines and metal systems is associated 
with galaxies (Bergeron et al. 1992; Lanzetta et al. 1995; 
Tytler 1995; Le Brune et al. 1996). However as was recently 
shown by Penton, Stock and Shull (2002), even at small 
redshifts some absorbers are associated with galaxy filaments 
while others are found within galaxy voids. These results 
suggest that the population of weaker absorbers dominating 
at higher redshifts can be associated with weaker structure 
elements formed by the non luminous baryonic and DM components 
in extended low density regions. In turn, the relatively 
homogeneous spatial distribution of absorbers implies a 
more homogeneous spatial distribution of both DM and 
baryonic components as compared with the observed 
distribution of the luminous matter.

For the truncated Zel'dovich theory (Coles, Melot\,\&\,
Shandarin 1993; Bond\,\&\,Wadsley 1997), some statistical
characteristics of absorbers were already discussed in 
Gnedin \& Hui (1996); Hui, Gnedin \& Zhang (1997); and Hui 
\& Rutledge (1999). In particular, in these papers possible 
fits for the number density of absorbers with a threshold 
hydrogen column density, $N_{HI}$, or with a threshold 
Doppler parameter, $b$, were proposed. However, because 
of the complex evolution of absorbers, these functions 
depend upon both the initial power spectrum and several 
random factors discussed below (Secs. 2.3, 4.3). So, the 
application and interpretation of these fits are 
limited. 

Here we use a statistical approach (DD02) based on the 
modified Zel'dovich theory which describes both linear and 
nonlinear evolution of the DM structure. For the CDM--like  
power spectrum and Gaussian initial perturbations it gives the 
expected mean 1D number density and the probability  distribution 
function of DM pancakes. Both functions  depend only on the 
cosmological model and moments of the initial power spectrum. 
Comparison of these functions with absorbers distributions 
confirms the self consistency of adopted physical model and 
Gaussianity of the initial density perturbations, allows to 
estimate the spectral moments and to restrict the mass of 
dominant DM particles. These results complement the 
investigations of the linear power spectrum (see, e.g., 
Croft, Weinberg, Katz \& Hernquist 1998; Gnedin \& Hamilton 
2002; Zaldarriaga, Scoccimorro \& Hui 2002; Croft et al. 2002). 

However, this theory does not describe the process of 
relaxation of compressed matter, the disruption of 
structure elements due to the gravitational instability 
of DM pancakes and the distribution of neutral hydrogen 
across DM pancakes. Now we have only limited information
about properties of the background gas and UV radiation   
(Scott et al. 2000; Schaye et al.  2000; McDonald \& 
Miralda--Escude 2001; McDonald et al. 2000, 2001; Theuns 
et al. 2002a, b). Therefore, in this paper some numerical 
factors remain undetermined. They could be more precisely 
determined by special investigations of simulated and 
observed absorbers. 

This analysis should be supplemented by application of the
discussed here approach to simulations which take 
simultaneously into account the impact of many important 
factors and provide unified picture of the process of 
absorbers formation and evolution (see, e.g., Weinberg 
et al. 1998; Zhang et al. 1998; Dav\'e et al. 1999; Theuns 
et al. 1999). But so far such simulations can be performed 
only in small boxes what restricts their representativity, 
introduces artificial cutoffs in the power spectrum and 
complicates the quantitative description of structure 
evolution (see more detailed discussion in Hui \& Gnedin  
1997 and Paper II). The semianalitic models (see, e.g., 
Bi \& Davidsen 1997; Choudhury, Padmanabhan \& Srianand 
2001; Choudhury, Srianand, \& Padmanabhan 2001; Viel et 
al. 2002) show other perspective approach to study the 
structure evolution at high redshifts. Perhaps, some results 
obtained below can be also useful for this purpose.
 
Comparison of results obtained in Paper I and Paper II 
and in this paper demonstrates that the quality and 
representativity of the sample of observed absorbers 
are very important for the reconstruction of processes 
of absorbers formation and evolution. In this paper we 
use the sample of $\sim$ 4500 absorbers at 1.7~$\leq~z\leq$~4 
compiled from 18 high resolution spectra. However, the redshift 
distribution of absorbers is inhomogeneous what increases 
errors and generates some uncertainties. In particular, the 
statistics of absorbers at $z\geq$~3.5 is very poor. Because 
of the preliminary character of our investigation we estimate 
the precision of only the most important parameters. 

Further progress can be achieved with richer samples covering 
the range of redshifts at least up to $z\sim$~4.5--5. The 
observational evidence of the reheating of the Universe at 
$z\approx$~6 (Djorgovski et al 2001; Fan et al. 2001) makes 
it important to perform a complex investigation of 
the early period  of structure evolution at redshifts 
$z\sim$~4--6.

This paper is organized as follows. The theoretical model 
of the structure evolution is discussed in Sec. 2. In Sec. 
3 the observational databases used in our analysis are 
presented. The results of statistical analysis are given 
in Secs. 4--6. Discussion and conclusion can be found in 
Sec. 7 .

\section{Model of absorbers formation and evolution}

The main observational characteristics of absorption lines 
are the redshift, $z_{abs}$, the column density of neutral 
hydrogen, $N_{HI}$, and the Doppler parameter, $b$, while 
the theoretical description of structure formation and 
evolution is dealing with the mean linear number density 
of absorbers, $n_{abs}(z)$, temperature, overdensity and 
entropy of DM and gaseous components, and with the 
ionization degree of hydrogen. To connect these theoretical 
and observed parameters a physical model of absorbers 
formation and evolution is required.

Many such models were proposed and discussed during the last 
twenty years (see references in Rauch 1998, and in Paper I 
and Paper II). However, to describe the properties of the new 
wider and refined sample of observed absorbers it is necessary 
to develop a more detailed physical model of absorbers as well. 
Our model includes some ideas discussed already in earlier 
publications.

\subsection{Physical model of absorbers.}

In this paper we assume that:
\begin{enumerate}
\item{} The DM distribution forms an interconnected structure 
of sheets (Zel'dovich pancakes) and filaments, their main 
parameters are approximately described by the Zel'dovich 
theory of gravitational instability applied to the CDM or WDM 
initial power spectrum (DD99; DD02). The majority of DM 
pancakes are partly relaxed, long-lived, and their 
properties vary due to the successive merging and expansion and/or
compression in the transverse directions.  
\item{} Gas is trapped in
the gravitational potential wells  formed by the DM distribution. The
gas temperature and the  observed Doppler parameter, $b$, trace the
depth of the DM  potential wells.
\item{} For a given temperature, the gas density within the
wells is determined by the gas entropy created during the 
previous evolution. The gas entropy is changing, mainly, 
due to shock heating in the course of merging of pancakes,  bulk
heating produced by the UV background and local  sources and due to
radiative cooling.  \item{} The gas is ionized by the UV background
and for  the majority of absorbers ionization equilibrium is assumed.
\item{} The observed properties of absorbers are changing 
because of merging, transversal compression and/or 
expansion and disruption of DM pancakes. The bulk heating 
and radiative cooling leads to slow drift of the entropy 
and density of the trapped gas. Random variations of the 
intensity and spectrum of the UV background enhance random 
scatter of the observed absorbers properties.
\item{} In this simple model we identify the velocity 
dispersion of DM component compressed within pancakes with 
the temperature of hydrogen and the Doppler parameter $b$ 
of absorbers. We consider the possible macroscopic motions 
within pancakes as subsonic and assume that they cannot 
essentially distort the measured Doppler parameter. 
\end{enumerate}
As compared with the model discussed in Paper II, here 
we take into account more accurately the evolution of 
pancakes after their formation.  

The formation of DM pancakes as an inevitable first step of 
evolution of small perturbations was firmly established both 
by theoretical considerations (Zel'dovich 1970; Shandarin \& 
Zel'dovich 1989) and numerical simulations (Shandarin et al. 
1995). Both theoretical analysis and simulations show the 
successive transformation of sheet-like elements into 
filamentary-like ones and, at the same time, the merging of 
both sheet-like and filamentary elements into richer walls. 
Such continuous transformation of structure goes on all the 
time. These processes imply the existence of a complicated 
time-dependent internal structure of high density elements 
and, in particular, the unavoidable arbitrariness in  discrimination
of such elements into filaments and sheets. 

Our approximate consideration cannot be applied to objects 
for which  the gravitational potential and the gas temperature 
along the line of sight strongly depends  on the matter 
distribution across this line. In this paper we use the term 
'pancake' to denote structure elements with relatively small 
gradient of properties along the line of sight. With such a 
criterion, the anisotropic halo of filaments and clouds can 
be also considered as 'pancake-like'. 

The subpopulation of weaker absorbers also contains 
"artificial" caustics (McGill 1990) and absorbers identified 
with slowly expanding underdense regions (Bi \& Davidsen 
1997; Zhang et al. 1998; Dav\'e et al. 1999). Such absorbers 
produce a short lived noise, which is stronger at higher 
redshifts $z\geq$~3. Fortunately, fraction of such absorbers 
does not exceed 10--15\% of the observed sample and they 
cannot significantly distort our final estimates. 

In this paper we consider the spatially flat $\Lambda$CDM 
model of the Universe with the Hubble parameter and mean 
density given by:
\be
H^{2}(z) = H_0^2\Omega_m(1+z)^3[1+\Omega_\Lambda/\Omega_m
(1+z)^{-3}]\,,
\label{basic}
\ee
\[
\langle\rho_m(z)\rangle = {3H_0^2\over 8\pi G}\Omega_m(1+z)^3,
\quad H_0=100h\,{\rm km/s/Mpc}\,.
\]
Here $\Omega_m=0.3\,\&\,\Omega_\Lambda=0.7$ are dimensionless 
matter density and the cosmological term, and $h=$~0.65 is 
the dimensionless Hubble constant.

\subsection{Properties of the homogeneously distributed 
hydrogen}

Properties of the compressed gas can be suitably related to 
the parameters of homogeneously distributed gas, which were 
discussed in many papers (see, e.g., Ikeuchi\,\&\,Ostriker  1986;
Hui \,\&\,Gnedin 1997; Scott et al. 2000; McDonald et  al. 2001;
Theuns et al. 2002a, b). Thus, the baryonic  density and
temperature can be taken as \be
\langle n_b(z)\rangle = 2.4\cdot 10^{-7}{\rm cm}^{-3}
(1+z)^3(\Omega_bh^2/0.02)\,,
\label{bg}
\ee
\[
T_{bg}\approx 1.6\cdot 10^4K\theta_{bg}^2(z),\quad 
b_{bg}=\sqrt{2k_BT_{bg}\over m_H}\approx 16\theta_{bg}
{\rm km/s}\,.
\]
Here $\Omega_b$ is the dimensionless mean density of 
baryons, $T_{bg}\,\&\,b_{bg}$ are the temperature and 
Doppler parameter of the gas, $\theta_{bg}^4(z)\sim 1+z$ 
describes a slow decrease of temperature with redshift  for low
density cosmological models, $k_B\,\&\,m_H$  are the Boltzmann's
constant and the mass of the  hydrogen atom.

Under the assumption of ionization equilibrium of the gas, 
\be
\alpha_rn_b^2\approx n_H\Gamma_\gamma,\quad 
\alpha_r(T)\approx 4\cdot 10^{-13}\left({10^4K\over T}
\right)^{3/4} {{\rm cm}^3\over{\rm  s}}\,,
\label{ieq}
\ee
where $\alpha_r(T)$ is the recombination coefficient (Black, 
1981) and $\Gamma_\gamma$ characterizes the rate of ionization 
by the UV background, the fraction of neutral hydrogen is
\be
x_{bg} = n_H/n_b = \langle n_b(z)\rangle~\alpha_r(T_{bg})
\Gamma_\gamma^{-1} = x_0(1+z)^3,
\label{xbg}
\ee
\[
x_0\approx {6.7\cdot 10^{-8}\over\Gamma_{12}}{\Omega_bh^2
\over 0.02}\left({16km/s\over b_{bg}(z)}\right)^{3/2},
\quad \Gamma_\gamma =10^{-12}s^{-1}\Gamma_{12}\,,
\]
For the standard power law spectrum of radiation we get 
\[ 
J(\nu)=J_{21}\cdot 10^{-21}\left({\nu_H\over\nu}\right
)^{\alpha_\gamma}erg~s^{-1}cm^{-2}sr^{-1}Hz^{-1}\,,
\]
\[
\Gamma_{12} =12.6J_{21}(3+\alpha_\gamma)^{-1}\,.
\]

The gas entropy can be characterized by the function
\be
F_{bg}={T_{bg}\over\langle n_b\rangle^{2/3}} = 
{36{\rm keV\cdot cm}^2\over (1+z)^2}\left({0.02\over\Omega_b
h^2}\right)^{2/3}\left({b_{bg}(z)\over 16km/s}\right)^2\,.
\label{sbg}
\ee

\subsection{Parameters of absorbers}

In this section we introduce the main relations between theoretical 
and observed characteristics of pancakes based on the physical 
model discussed in Sec. 2.1. The most important characteristics 
of DM pancakes are presented (without proofs) as a basis for 
further analysis. For more details see DD99 and  DD02.

\subsubsection{Characteristics of DM component}

The fundamental characteristic of DM pancakes is the 
dimensional, $\mu$, or the dimensionless, $q$, Lagrangian 
thickness (the dimensionless DM column density) :
\be
\mu\approx {\langle\rho_m(z)\rangle l_vq\over (1+z)} = 
{3H_0^2\over 8\pi G}l_v\Omega_m(1+z)^2q\,, 
\label{mu}
\ee
\[
l_v\approx {6.6\over h\Omega_m}h^{-1}{\rm Mpc}=33.8 h^{-1}
{\rm Mpc}{0.3\over\Omega_m}{0.65\over h}\,,
\]
where $l_v$ is the coherent length of initial velocity field 
(DD99; DD02). The Lagrangian thickness of a pancake, $l_v q$, 
is defined as the unperturbed distance at redshift $z=0$
between  DM particles bounding the pancake. The actual 
thickness of a pancake is 
\be \Delta r = \mu\rho_m^{-1} =
l_vq(1+z)^{-1}\delta^{-1}, \quad
\delta=\rho_m/\langle\rho_m(z)\rangle\,.  \label{delr}
\ee
Here $\delta$ is the mean overdensity of a pancake above 
the background density. 

Comparing the pancake surface density, $\mu(z)$, with the 
distance between neighboring absorbers,  
\be
D_{sep}={c\Delta z\over H(z)}=5.5\cdot 10^3{\Delta z\over 
(1+z)^{3/2} }\sqrt{0.3\over\Omega_m}h^{-1}{\rm Mpc}\,,
\label{sep}
\ee
we can also estimate the fraction of matter accumulated by 
absorbers as
\be
f_{abs}\simeq{\mu(1+z)\over\langle\rho_m\rangle D_{sep}}
\approx 6.2\cdot 10^{-3}q{(1+z)^{3/2}\over\Delta z}
\sqrt{0.13\over\Omega_mh^2}\,.
\label{fract}
\ee
The fraction of clustered matter characterized by $\langle
f_{abs}(z)\rangle$ can be compared with theoretical  expectations
(see Sec. 4.4). 

For Gaussian initial perturbations, the expected probability 
distribution function for the DM column density is
\be
N_qdq\approx {2\over\sqrt{\pi}} e^{-\xi}~{{\rm erf}
(\sqrt{\xi})\over\sqrt{\xi}}\left({q^2\over q^2+q_0^2}\right
)^{3/4}d\xi,\quad \xi={q\over 8\tau^2}\,,
\label{qpdf}
\ee
\[
\langle\xi\rangle\approx{1\over 2}+{1\over\pi}\approx 0.82,\quad 
\langle\xi^2\rangle\approx{3\over 4}+{2\over\pi},\quad 
\langle\xi^3\rangle\approx{3\over 16}+{11\over 2\pi}\,,
\]
(DD02), where the parameter $q_0\ll$ 1 characterizes the 
coherent length of the initial density field (see Sec. 7.5) and 
the dimensionless 'time' $\tau(z)$ describes the evolution 
of perturbations in the Zel'dovich theory. For $\Lambda$CDM 
model (\ref{basic}) and $z\geq 2$ we have
\be
\tau(z)\approx \tau_0\left({1+1.2\Omega_m\over 2.2\Omega_m}
\right)^{1/3}{1\over 1+z}\approx {1.27\tau_0\over 1+z}\,,
\label{Btau}
\ee
and for the amplitude of perturbations (Doroshkevich, Tucker 
\& Allam 2002)
\be
\tau_0\approx (0.27\pm 0.04)\sqrt{\Omega_mh\over 0.2},\quad 
\xi\approx 1.06q(1+z)^2{ 0.2\over\Omega_mh}\,.
\label{tau0}
\ee

Strictly speaking, relations (\ref{mu}) and (\ref{qpdf}) are 
valid for pancakes formed and observed at the same redshift 
$z_{obs}=z_f$ and after pancake formation the transverse 
expansion and compression changes its DM column density. 
However, as was shown in DD02, these processes do not change 
the statistical characteristics of pancakes observed at 
redshift $z_{obs}\leq z_f$. This means that statistically 
we can consider each pancake as created at the observed 
redshift. More details are given in DD99 and DD02.

For the DM dominated cosmological model the observed Doppler 
parameter is also closely linked with properties of the DM 
pancakes. Thus, in Paper II the $b$--parameter was identified 
with the infall velocity of matter into pancakes:
\be
b\approx v_{inf}=u_0(z)\sqrt{\xi^2+2\xi}\,, 
\label{vinf}
\ee
\[
u_0(z)\approx 2.3l_vH_0\tau^2\sqrt{1+z}\approx {310 km/s\over 
(1+z)^{3/2}}\sqrt{0.13\over\Omega_mh^2}\,.
\]
This assumption is valid also some time after formation of the
pancake but later on the pancake relaxation, compression 
and/or expansion in transverse directions change the 
$b$--parameter. 

As is seen from (\ref{vinf}), for $\langle\xi\rangle$ 
independent from redshift (\ref{qpdf}), a systematic variation 
of the mean Doppler parameter with redshift could be expected. 
The problem was left open in Paper II (due to limited 
observational data set) but now, with the more representative 
sample of absorbers, we see surprisingly weak redshift 
variations of the mean observed Doppler parameter (Sec. 4.1). 
This means that only small fraction of observed absorbers 
is 'young' and is described by (\ref{vinf}), whereas 'older' 
relaxed and gravitationally confined absorbers dominate this 
sample. The simple model of relaxed absorbers was already 
considered in Paper II. Here we substantially improve it.  

As is well known, for an equilibrium slab of DM the depth of
potential well is
\be
\Delta \Phi\approx {\pi G\mu^2\over \langle \rho(z)\rangle
\delta}\Theta_q^2\Theta_\Phi\,,
\label{phi}
\ee
where random factors $\Theta_\Phi\,\&\,\Theta_q$ 
characterizes the nonhomogeneity of DM distribution across 
the slab and the evaporation of matter in the course of its 
relaxation. 

Analysis of numerical simulations (Demia\'nski et al. 2000) 
indicates that the relaxed distribution of DM component can be 
approximately described by the politropic equation of state 
with the power index $\gamma\approx$ 2\,. In this case, we  can
expect that  
\[
\Theta_\Phi = {2\gamma\over\sqrt{\pi}(\gamma-1)}
{\Gamma(1.5-1/\gamma)\over\Gamma(1-1/\gamma)}= {4\over\pi} \,.
\]
where $\Gamma(y)$ is Euler function.
The actual distribution of DM component across a slab and 
the value of $\Theta_\Phi$ depend upon the relaxation process 
which is essentially accelerated due to the pancake disruption 
into the system of high density clouds and filaments. This 
means that the factor $\Theta_\Phi$ randomly vary from absorber 
to absorber. 

At each pancake merging $\sim$ 10 -- 15\% of matter is evaporated 
due to the violent relaxation. This means that $\Theta_q\sim$ 1 
for absorbers formed from the homogeneously distributed matter 
and $\Theta_q$ decreases for rich absorbers formed due to several 
successive  mergings.

The Doppler parameter is defined by the depth of potential 
well and for the isentropic gas with $p_{gas}\propto 
\rho_{gas}^{5/3}$ trapped within the well, we get
\be
b^2\approx {4\over 5}\Delta \Phi\approx {4\over 5}{\pi 
G\mu^2\over\langle \rho(z)\rangle\delta}\Theta_q^2
\Theta_\Phi=\delta_0b_{bg}^2{q^2\over\delta}(1+z)\,,
\label{bb}
\ee
\[
\delta_0 = {3\over 10}\left({H_0l_v\over b_{bg}}\right)^2
\Omega_m\Theta_\Phi\Theta_q^2 \approx 4\cdot 10^3\Theta_
\delta\,,
\]
\[
\Theta_\delta = \left({0.13\over\Omega_m h^2}\right)
\left({16km/s\over b_{bg}(z)}\right)^2\Theta_\Phi\Theta_q^2\,.
\]
Variations of the gas entropy across an absorber increase the
random variations of $\Theta_\delta$ and $b$.

\subsubsection{Characteristics of gaseous component}

The observed column density of neutral hydrogen can be written
as an integral over a pancake along the line of sight
\be
N_{HI} = \int dx ~\rho_b x_H = 2\langle x_H\rangle{\langle n_b
(z)\rangle l_v q\over 1+z}{0.5\over cos\theta}\,.
\label{NH1}
\ee
Here $\langle x_H\rangle$ is the mean fraction of neutral 
hydrogen and $cos\theta$ takes into account the random 
orientation of absorbers and the line of sight ($\langle 
cos\theta\rangle\approx 0.5)$. We assume also that both DM 
and gaseous components are compressed together and, so, the 
column density of baryons and DM component are proportional 
to each other. 

However, the overdensity of the baryonic component, $\delta_b 
= n_b/\langle n_b\rangle$ is not identical to the overdensity 
of DM component, $\delta$, (see, e.g., discussion in Matarrese 
\& Mohayaee 2002). Indeed, the gas temperature and the Doppler 
parameter are mainly determined by the characteristics of DM 
component (\ref{bb}) but the gas overdensity is smaller than 
that of DM component due to larger entropy of the gas. Moreover, 
the bulk heating and cooling change the density and entropy of 
the gas trapped within the DM potential well. Under the 
assumption of ionization equilibrium of the gas (\ref{ieq}) 
this process is described by the equation
\be
{1\over n_b}{dn_b\over dt} = n_b\alpha_r(T)[\varepsilon(T)-
T_\gamma/T],\quad T=T_{bg}b^2/b_{bg}^2\,,
\label{thermal}
\ee
where the recombination coefficient $\alpha_r(T)$ was 
given in (\ref{xbg}), $T_\gamma\sim (5 - 10)\cdot 10^4$K 
characterizes the energy injected at a photoionization and 
\[
\varepsilon(T)\approx 2T_4^{1/4}[1-0.3\ln T_4+0.13T_4^{1/3}],
\quad T_4=T/10^4K\,,
\] 
describes the radiative cooling for the bremsstrahlung emission 
and recombination of hydrogen and helium (Black 1981). 

The energy injected at the photoionization, $T_\gamma$, 
depends upon the spectrum of local UV background and it 
varies randomly with  time and space. The gas temperature 
varies because of the pancake expansion and compression and 
due to the pancake disruption into a system of high density 
clouds. As is seen from (\ref{thermal}), these processes 
change the baryonic density of pancakes and we  can write \be
\delta_b = \kappa_b(z)\delta\,.  \label{kappab}
\ee 
This means that the factor $\kappa_b$ is small for absorbers 
formed due to adiabatic and weak shock compression because 
of the large difference between entropies of the background DM 
and the gas, and $\kappa_b\rightarrow$ 1 for richer hot absorbers 
formed due to strong shock compression when this difference 
becomes small. 

Under the  assumption of ionization equilibrium of the gas 
(\ref{ieq}) and neglecting a possible contribution of 
macroscopic motions to the $b$-parameter ($T\propto b^2$), 
for the fraction of neutral hydrogen and its column  density we get:
\[ \langle x_H\rangle = x_0\kappa_b(z)\delta\beta^{-3/2}(1+z)^3
\Theta_x,\quad \beta=b/b_{bg}\,,
\]
\be
N_{HI} = N_0q\delta\beta^{-3/2}(1+z)^5,\quad 
N_0 = 5\cdot 10^{12}cm^{-2}\Theta_H\,,
\label{NH2}     
\ee
\[
\Theta_H={\kappa_b\Theta_x\over\Gamma_{12}}{0.13\over
\Omega_mh^2}{\langle cos\theta\rangle\over cos\theta}
\left({\Omega_bh^2\over 0.02}\right)^2\left({16km/s\over 
b_{bg}(z)}\right)^{3/2}\,,
\] 
where $\Gamma_{12}$, $b_{bg}$ and $x_0$ were defined in 
(\ref{bg}) and (\ref{xbg}) and the factor $\Theta_x$ describes  the
nonhomogeneous distribution of ionized hydrogen along  the line of
sight.  

\subsubsection{Absorbers characteristics}

Eqs. (\ref{bb}) and (\ref{NH2}) link the observed and  
other physical characteristics of absorbers. For the DM 
column density, $q$, the average gas entropy, $\Sigma=
\ln(F_s)$, and overdensity, $\delta$, we get:
\be
q^3 = {N_{HI}\over N_0\delta_0}{\beta^{7/2}
\over (1+z)^6},\quad \delta = \delta_0{q^2\over\beta^2}(1+z)\,,
\label{main}
\ee
\[
\exp(\Sigma)=F_s = \beta^2/\delta_b^{2/3}=\kappa_b^{-2/3}
\beta^2/\delta^{2/3}\,.
\]
Here we use the standard equation of state $T/T_{bg}=\beta^2
(z) = F_s(z)\delta^{2/3}(z)$ . 

The precision of these estimates is moderate and the main 
uncertainties are generated by the unknown cos$\theta$ and 
parameters $\Theta_\Phi$, $\Theta_q$, $\Theta_x$, $\kappa_b$ 
and $\Gamma_\gamma$, which vary -- randomly and 
systematically -- from absorber to absorber (estimates of 
$q$ and $\delta$ are independent from $b_{bg}$). These 
variations distort parameters defined in (\ref{main}) as 
follows:
\be
q^3\propto {1\over\Theta_H\Theta_\delta},\quad \delta^3
\propto{\Theta_\delta\over\Theta_H^2},\quad F_s\propto
\kappa_b^{-2/3}\left({\Theta_H^2\over\Theta_\delta}\right
)^{2/9}\,.
\label{gam}
\ee
Independent estimates of these uncertainties can be obtained 
from the analysis of the redshift distribution of absorbers 
(Sec. 6). 

However, comparing the average DM column density of pancakes,
$\langle q\rangle$, with expectations (\ref{qpdf}) we can  
restrict the possible redshift variations of $\langle\Theta_H
\Theta_\delta\rangle$ and estimate the combined uncertainty 
introduced by unknown factors. As is seen from (\ref{main}), 
\[
\xi^3\approx (1+z)^6~q^2 = 
\left({N_{HI}\over 2\cdot 10^{16}cm^{-2}}\right)
\left({b\over 16km/s}\right)^{7/2}G_{12}\,,
\]
\be
G_{12}(z)={\Gamma_{12}\over \kappa_b\Theta_x\Theta_\Phi
\Theta_q^2}{cos\theta\over 0.5}\left({\Omega_mh^2
\over 0.13}{\Omega_bh^2\over 0.02}\right)^2\,.
\label{g12}
\ee
This relation shows that, for statistically homogeneous 
sample of absorbers with $\langle\xi^3\rangle =const.$, 
$\langle cos\theta\rangle=$~0.5,
we can estimate the redshift variations of $\langle G_{12}(z)
\rangle$ as follows:
\[
\langle G_{12}(z)\rangle\propto\left\langle{\Gamma_{12}\over 
\kappa_b\Theta_x\Theta_\Phi\Theta_q^2}\right\rangle\propto
\langle\xi^3\rangle/\langle N_{HI} b^{7/2}\rangle\,.
\]
Precision of these estimates is limited because of strong 
random scatter of the product $N_{HI} b^{7/2}$ . 

\subsubsection{Regular and random variations of absorber 
characteristics}

The most fundamental characteristic of absorbers is their 
DM column density, $\xi\approx q(1+z)^2$. It describes the 
formation and merging of pancakes, is only weakly sensitive to  the
action of random factors and defines the regular  variations of the
Doppler parameter, $b$, overdensity,  $\delta$, and entropy, $F_s$.
These parameters include  also a random component which integrates
the evolutionary history of each pancake and the action of random
factors discussed in the previous subsection. If the structure of 
a relaxed pancake can be described by the politropic 
equation of state with the effective power index $\gamma$ 
then we can discriminate the regular and random variations 
of $b,\,\delta,\,\&\,F_s$ and introduce the {\it reduced} 
characteristics of relaxed absorbers, $\upsilon,\,\Delta\,
\&\,S$:
\[
\upsilon = \ln[\beta\xi^{(1/\gamma-1)}]\,, 
\]
\be
\Delta = \ln[(1+z)^3\delta/\xi^{2/\gamma}] = 
\ln(\delta_0)-2 \upsilon\,,
\label{reduct}
\ee
\[
S = \ln[(1+z)^{-2}F_s\xi^{2(5/3-\gamma)/\gamma}] =
const. + 10 \upsilon/3\,.
\]
These relations indicate that, in fact, our approximate 
description use only {\it one} random characteristic, namely, 
$\upsilon$, which is expressed through observed parameters 
as follows: 
\be
\upsilon = {7-\gamma\over 6\gamma}\ln~\beta-
{\gamma-1\over 3\gamma}\ln\left({N_{HI}\over N_0\delta_0}\right)\,.
\label{nbh}
\ee 

As is shown in Secs. 4.5.2, the choice of a suitable 
value of $\gamma$ allows to minimize the correlation between 
$\xi$ and $\upsilon$. Of course, this reduction is approximate 
and it can be achieved only statistically for a given sample  of
absorbers. 

\subsection{Mean number density of absorbers}

Following  Paper I we will characterize the 1D mean  number density
of absorbers by the dimensionless function  \be
n_{abs}={c\over H_0}\langle l\rangle^{-1} = {H(z)\over H_0}
{dN(z)\over dz}\,.
\label{nabs}
\ee
Here $dN(z)$ is the mean number of absorbers between $z$ and 
$z+dz$, $\langle l(z)\rangle$ is the mean free path between 
absorbers at redshift $z$ and $c$ is the speed of light. When 
absorbers are identified with the Zel'dovich pancakes, this 
1D mean number density can be linked with the fundamental 
characteristics of the cosmological model and moments of  the
initial power spectrum (DD02 \& Sec. 7.5). 

As was shown in DD02, for Gaussian initial perturbations and 
for richer DM pancakes with sizes significantly larger than 
the coherent length of initial density field, this number 
density can be approximated by the expression 
\be
n_{abs}\approx {c\over H_0l_v}{W_p(\xi_{thr})(1+z)^2\over
\langle q(\xi_{thr})\rangle}\,,
\label{nw}
\ee
\[
\langle q(\xi_{thr})\rangle=4\tau^2\left[1+{4\sqrt{\pi\xi_{thr}}
\mbox{erf}(\sqrt{\xi_{thr}})+2\exp(-\xi_{thr})\over \pi\exp(
\xi_{thr})[1-\mbox{erf}^2(\sqrt{\xi_{thr}})]}\right]\,,
\]
\[
W_p(\xi_{thr})\approx 0.5[1-\mbox{erf}(\sqrt{\xi_{thr}})]\,,
\]
where $\xi_{thr}(z)=q_{thr}/8\tau^2(z)$, $q_{thr}(z)$ 
is the minimal (threshold) DM column density of pancakes, 
$W_p(\xi_{thr})$ and $\langle q(\xi_{thr})
\rangle$ are the fraction of matter and the mean DM column 
density for pancakes with $q\geq q_{thr}$ and the factor  $(1+z)^2$
describes the cosmological expansion of pancakes  population.

The expression (\ref{nw}) contains only one fitting 
parameter, $\xi_{thr}$, and it describes quite well the 
evolution of richer pancakes. However, application of this 
relation to evolution of weaker pancakes is problematic as 
for $\xi_{thr}\ll$ 1 we get from (\ref{nw}):
\[
n_{abs}\approx 55(1+z)^4\,, 
\]
and this relation does not contain any fitting parameters.

The formation of pancakes with small $q$ is suppressed 
due to the small scale cutoff in the initial power spectrum. 
For pancakes with small $q_{thr}$ the redshift variations of 
the mean number density is described by the expression
\be
n_{abs}\approx {c\over H_0l_v}{\sqrt{3}(1+z)^2\over 
16\pi\tau(z)\sqrt{q_0}}{\mbox{erf}(\sqrt{\xi_{thr}})\over
\sqrt{\xi_{thr}} }\exp(-\xi_{thr})\,,
\label{nq0}
\ee
(DD02) where again $\xi_{thr}(z)=q_{thr}(z)/8\tau^2$ and 
the value $q_0$ characterizes the coherent length of the 
initial density field (see Sec. 7.5). 

Both expressions, (\ref{nw}) and (\ref{nq0}), are derived for the
Gaussian initial perturbations. They are valid  for DM pancakes
formed due to both linear and nonlinear  compressions and allow for
merging of pancakes. Moreover,  as was shown in DD02, if the
observed redshift is identified  with the redshift of pancake
formation then both relations  remain the same even when the
transverse compression and  expansion of pancakes are also taken
into account. This means that they partly account for the pancake
disruption as well. 

The relation (\ref{nq0}) describes also the  impact of the gaseous
pressure on the formation of observed  absorbers. In this case, the
parameter $q_0$ must be calculated with the power spectrum corrected
for the Jeans damping of  small scale perturbations in the baryonic
component  (see Sec. 7.5). 

\subsection{Observational restrictions}

The completeness of the observed samples of absorbers is 
restricted by the condition $N_{HI}\geq N_{thr}\approx 
10^{12} cm^{-2}$ what in turn distorts the characteristics 
of observed absorbers and makes it difficult to compare 
them with the theoretical expectations discussed in Secs. 
2.3 and 2.4 . This means that these expectations should 
be corrected for the impact of the threshold column 
density of observed absorbers. 

Thus the investigation of spatial distribution of galaxies in 
the SDSS EDR (Doroshkevich, Tucker \& Allam 2002) results in 
estimates of typical parameters of galaxy walls at small 
redshifts as
\be
\langle q\rangle \approx 0.4,\quad \langle\beta\rangle\approx 
20,\quad \langle\delta\rangle\approx 3\,.
\label{walls}
\ee
With these data the expected column density of 
neutral hydrogen within the typical wall (\ref{NH1}) is
\[
N_{HI}\approx 0.02 N_0\approx 0.1 N_{thr}\Theta_H\leq N_{thr}\,.
\]
This means that even so spectacular object as the 'Greet Wall' 
does not manifest itself as absorbers. Rare absorbers with 
$b\geq$ 100 km/s, $N_{HI}\sim 10^{13} cm^{-2}$ and $\xi
\sim$~1.5 observed at $z\sim$ 1.5 -- 4 can be associated with 
embryos of wall--like elements of the Large Scale Structure of 
the Universe (see Sec. 5 for more detailed discussion). 

For absorbers associated with 'young' pancakes the Doppler 
parameter is given by (\ref{vinf}) and for the richer 
'young' pancakes with the mean DM column density we get 
\[
\xi\approx 1,\quad q\approx \langle q\rangle\approx (1+z)^{-2}, 
\quad \beta\approx 35(1+z)^{-3/2}\,,
\]
\be
\delta\approx 3~\Theta_\delta,\quad N_{HI}\approx 10^{-2}
(1+z)^{21/4}N_0\Theta_\delta\,,
\label{nhmn}
\ee
and such pancakes can be seen as absorbers already at redshifts 
$z\geq$ 1.

For more numerous poorer pancakes with $q\leq \langle q\rangle $ 
we have 
\be
\beta\sim 30\sqrt{q}(1+z)^{-1/2},\quad \delta\approx 
4q(1+z)^2\Theta_\delta\,,
\label{poor}
\ee
\[
 N_{HI}\approx 2.5\cdot 10^{-2}N_0
q^{5/4}(1+z)^{31/4}\Theta_\delta
\]
\[
\approx \left({q\over 0.02}\right)^{5/4}\left({1+z\over 2.5}
\right)^{31/4}N_{thr}\Theta_\delta\Theta_H\,,
\]
and such pancakes become visible for $z\geq$ 1.5 only. 

These rough estimates of expected characteristics of 
absorbers show that: 
\begin{enumerate}
\item    At redshifts $z\leq$ 1.5 we can observe mainly old 
pancakes formed at higher redshifts which kept the measurable 
column density of neutral hydrogen up to small redshifts. 
\item  The expected number density of observed absorbers 
rapidly increases at redshifts $z\sim$ 1.5 -- 2 when even 
relatively poor 'young' pancakes can pass over the 
observational threshold and become seen as weak absorbers. 
\end{enumerate}

This means that our analysis can be applied to absorbers 
observed at $z\geq$ 1.5 -- 2 when the impact of the 
observational threshold becomes moderate. 

\subsection{Variations of the UV background}

Direct estimates of the intensity of UV background through 
the proximity effect (Scott et al. 2000) give $\Gamma_{12}
\approx 2\pm 1$ at $z\sim$  2.5 -- 3 while McDonald \& 
Miralda-Escude (2001) got $\Gamma_{12}\sim 0.7 - 0.4$ at 
the same $z$. Indirect estimates of $\Gamma_{12}$ (Songaila 
1998) demonstrate a probable sudden drop of the UV intensity 
of about 3 times at $z\sim$ 3. This effect can be related to 
the strong ionization of HeII at these redshifts (Jacobsen 
et al. 1994; Zeng, Davidsen \& Kriss 1998; Theuns at al. 
2002a, b). 

As is shown in Sec. 6.2, possible redshift variations of 
$\Gamma_{12}$ are also seen in the redshift distribution 
of weaker absorbers at $z\approx$ 2.5. These variations 
can be fitted by the expression
\be
\Gamma_{12} = {a_0+1\over 2}\left[1+{a_0-1\over a_0+1}
{\rm th}\left({z-z_\gamma\over 0.16}\right)\right]\,,
\label{zgamma}
\ee
where $a_0\sim$ 1.5 -- 3 and $z_\gamma\sim$ 2.5 -- 3 
characterize the amplitude and the redshift of these 
variations. These estimates can be distorted due to the 
limited representativity of our sample and a  
nonhomogeneous redshift distribution of observed absorbers 
plotted in Fig. 1 . 

However, weak redshift variations of the functions $\langle 
b\rangle,$ $\langle\lg(N_{HI})\rangle\,\&\,\langle G_{12}
\rangle$ (Sec. 4.1) show that the influence of variations 
of $\Gamma_{12}$ on the properties of absorbers is partly 
compensated by variations of $\kappa_b$, $\Theta_x$, 
$\Theta_q\,\&\,\Theta_\Phi$ introduced in Secs. 2.3.1\,\&
\,2.3.2\,. As is found in Sec. 4.2, a reasonable 
description of the mean absorbers characteristics is achieved 
when the function $\langle G_{12}\rangle$ (\ref{g12}) is  
\be
\langle G_{12}\rangle\approx G_0(1-0.17 z),\quad G_0=50\,.
\label{g12z}
\ee

Comparison of this function with estimates of $\Gamma_{12}$ 
(Scott et al. 2000; McDonald \& Miralda-Escude 2001) allows 
to estimate the unknown factors as
\[
\langle\kappa_b\Theta_x\Theta_\Phi\Theta_q^2\rangle\sim 
(0.1 - 0.05)\left({\Omega_mh^2\over 0.13}{\Omega_b
h^2\over 0.02}\right)^2\,.
\]  
More accurate estimate of these factors can be obtained  
with numerical simulations. 

\begin{table}
\caption{QSO spectra used}
\label{tbl1}
\begin{tabular}{ccccr}
    &$z_{em}$&$z_{min}$&$z_{max}$&No \\
               &     &   &   &of HI lines\\
$0000-260^{1}~$ & 4.11&3.4&4.1& 431\\   
$0055-259^{2}~$ & 3.66&3.0&3.6& 534\\       
$0014+813^{3}~$ & 3.41&2.7&3.2& 262\\   
$0956+122^{3}~$ & 3.30&2.6&3.1& 256\\       
$0302-003^{3, 2}$&3.29&2.6&3.1& 356\\   
$0636+680^{3}~$ & 3.17&2.5&3.0& 313\\       
$1759+754^{4}~$ & 3.05&2.4&3.0& 307\\       
$1946+766^{5}~$ & 3.02&2.4&3.0& 461\\       
$1347-246^{2}~$ & 2.63&2.1&2.6& 361\\       
$1122-441^{2}~$ & 2.42&1.9&2.4& 353\\       
$2217-282^{2}~$ & 2.41&1.9&2.3& 262\\       
$2233-606^{6}~$ & 2.24&1.5&2.2& 293\\       
$1101-264^{2}~$ & 2.15&1.6&2.1& 277\\       
$0515-441^{2}~$ & 1.72&1.5&1.7&~~76\\       
\hline
$2126-158^{7}~$ & 3.26&2.9&3.2& 130\\   
$1700+642^{8}~$ & 2.72&2.1&2.7&~~85\\   
$1225+317^{9}~$ & 2.20&1.7&2.2& 159\\   
$1331+170^{10}~$& 2.10&1.7&2.1&~~69\\   
\vspace{0.15cm}
\end{tabular}                                   

1. Lu et al. (1996),
2. unpublished, courtesy of Dr. Kim   
3. Hu et al., (1995),
4. Djorgovski et al. (2001)
5. Kirkman \& Tytler (1997),
6. Cristiani \& D'Odorico (2000),
7. Giallongo et al. (1993),
8. Rodriguez et al. (1995),
9. Khare et al. (1997),
10. Kulkarni et al. (1996).
\end{table}

\section{ The database.}

The present analysis is based on the 18 spectra listed in 
Table 1. The available Ly-$\alpha$ lines were arranged into 
three samples. The richest sample, $S_{14}^{12}$, includes 
4369 absorbers with $10^{15}cm^{-2}\geq N_{HI}\geq 10^{12}
cm^{-2}$ from the first 14 high resolution spectra. It 
can be partly incomplete. For comparison, we use the most 
reliable sample $S_{18}^{13}$ which includes 2643 absorbers 
from all 18 spectra for $10^{13}cm^{-2}\leq N_{HI}\leq 10^{
15}cm^{-2}$ . These lines are more easily identified 
and they are not so sensitive to outer random influences. 
The sample $W_{14}^{12}$ contains 2126 weaker lines with 
$N_{HI}\leq 10^{13}cm^{-2}$ from the first 14 QSOs. It is 
mainly used to characterize possible variations of the UV 
background. To decrease the scatter of absorbers 
characteristics we exclude from these samples 45 absorbers 
with $b\geq$ 90km/s\,.

As is seen from Fig 1, for all samples the redshift 
distribution of absorbers is nonhomogeneous and the 
majority of absorbers are concentrated at 2~$\leq z
\leq$~3\,. This means that some of the discussed absorbers 
characteristics are derived mainly from this range of redshifts. 
Absorbers at $z\geq~3.5$ were identified mainly from the 
spectrum of QSO 0000-260 (Lu et al. 1996) and here the 
line statistics is insufficient. 
 
\begin{figure}
\centering
\epsfxsize=7.cm
\epsfbox{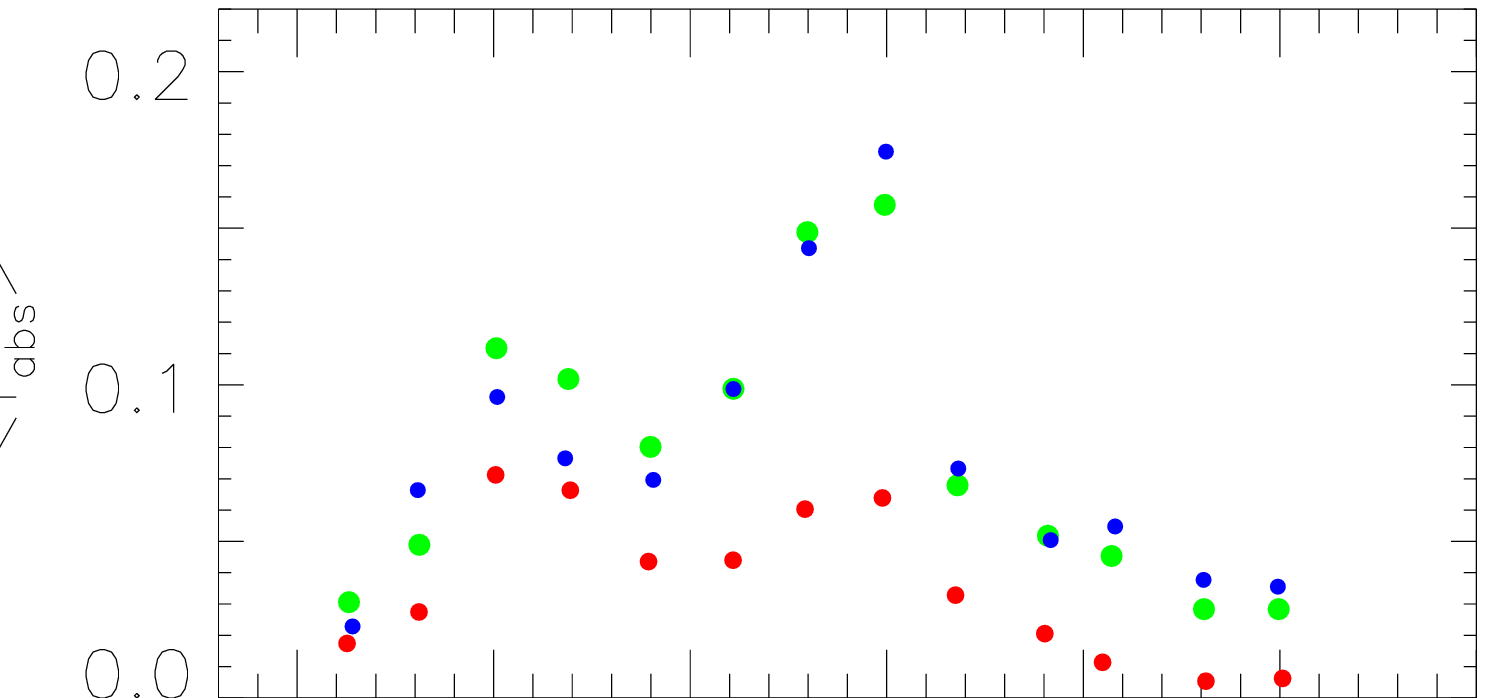}
\vspace{1.cm}
\caption{Redshift distribution of the fraction of observed 
absorbers for samples $W_{14}^{12}$ (red points), 
$S_{14}^{12}$ (green points) and $S_{18}^{13}$ (blue points). 
} 
\end{figure}

\section{Statistical characteristics of absorbers}

In this Section, the functions $q, \xi, \delta~\&~F_s$ are 
found for the $\Lambda$CDM cosmological model (\ref{basic}) 
with $b_{bg}=$~16km/s, $\Omega_bh^2=$~0.02 and the function 
$\langle G_{12}(z)\rangle$ given by (\ref{g12z}) with  
\be
G_0=50,\,\Theta_\delta= \langle\Theta_\Phi\Theta_q^2
\rangle=0.5,\, \Theta_H=\left\langle{\kappa_b\Theta_x
\over\Gamma_{12}}\right\rangle=0.1. 
\label{cosm}
\ee
Some quantitative results of this Section depend on the 
completeness and representativity of the samples and the 
quite arbitrary choice of $\Theta_\delta\,\&\,\Theta_H$ 
(\ref{cosm}) through the relations (\ref{gam}). As was 
noted above, for each absorber these factors change in 
the course of absorbers evolution and, in fact, the 
accepted values are averaged over the sample. For the 
sample $S_{14}^{12}$,~ $\sim$~10-15\% of the weaker 
absorbers could be related to the artificial "noise". 

For majority of QSO spectra used in our analysis the scatter 
of observed $N_{HI}$ is less then 10 per cents, while the 
scatter of the measured Doppler parameters, $b$, can achieve 
20 --30 per cents. Non the less, dispersions of absorbers 
characteristics discussed below are defined mainly by 
their broad distribution functions and by the completeness 
of the samples. Because of this, we discuss in this Section 
uncertainties of only the more interesting quantitative 
characteristics of absorbers. 
 
\begin{figure}
\centering
\epsfxsize=7.cm
\epsfbox{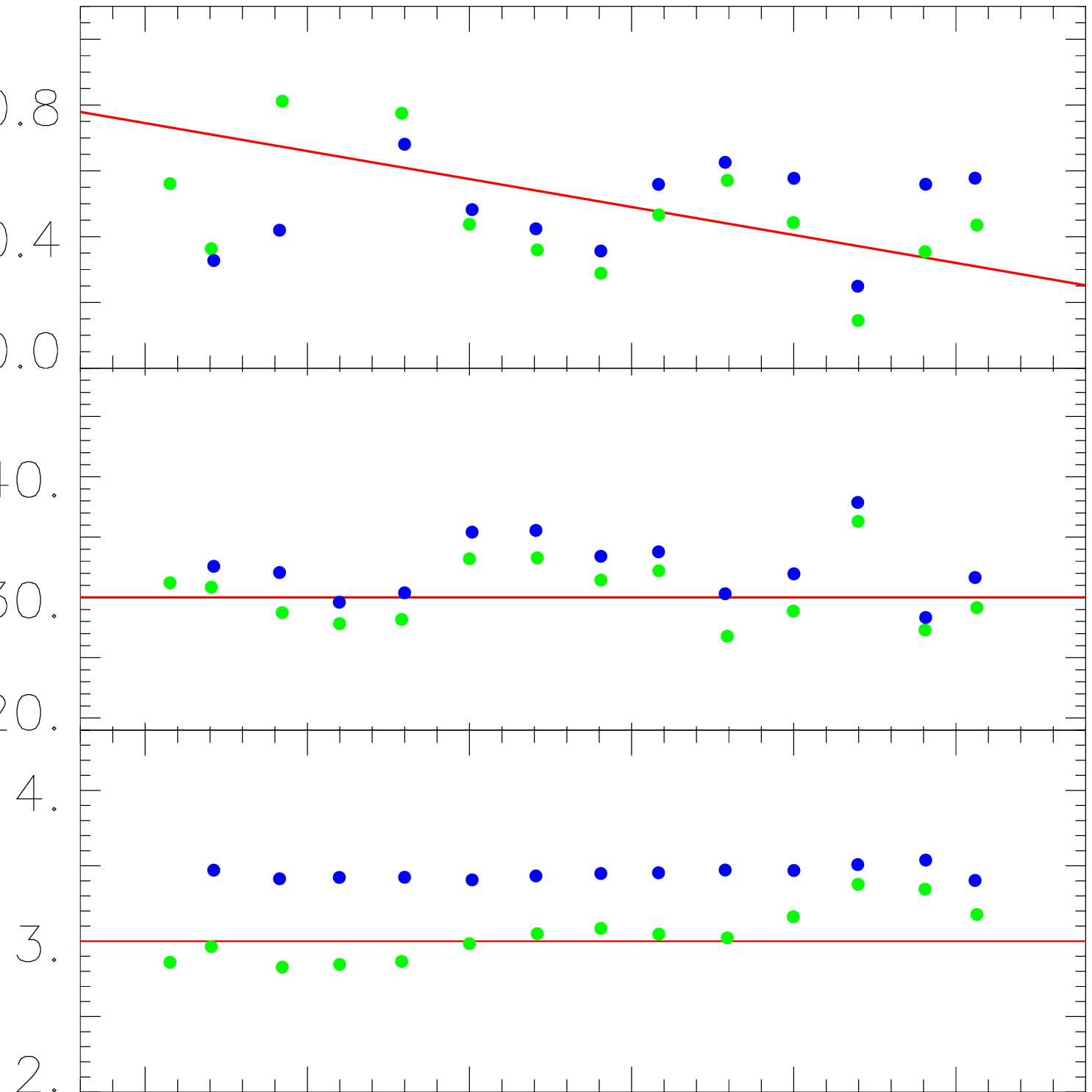}
\vspace{1.cm}
\caption{Redshift variations of $\langle G_{12}
\rangle$ (top panel), Doppler parameter, $\langle 
b\rangle$ (middle panel) and $\langle lg(N_{HI})\rangle$ 
(bottom panel) for samples $S^{13}_{18}$ (blue points) 
and $S^{12}_{14}$ (green points). Mean values (\ref{mnsbh}) 
and fit (\ref{g12z}) are plotted by straight lines.
} 
\end{figure}

\subsection{Redshift variations of the mean observed 
characteristics}

Redshift variations of two observed characteristics of 
absorbers, $\langle b\rangle$, and $\langle lg N_{HI}
\rangle$, are plotted in Fig. 2. For both samples $S_{14}
^{12}$ and $S_{18}^{13}$, the mean values $\langle 
b\rangle$ and $\langle lg N_{HI}\rangle$  
surprisingly weakly vary with redshift,
\be
\langle b\rangle = (28.6\pm 1.6){\rm km}/{\rm s},\quad \langle 
b\rangle = (31.7\pm 1.6){\rm km}/{\rm s}, 
\label{mnsbh}
\ee
\[
\langle lg N_{HI}\rangle = (13.1\pm 0.22),\quad \langle 
lg N_{HI}\rangle = (13.6\pm 0.06), 
\]
respectively. At the same time, the mean Doppler parameter 
systematically shifts from $\langle b\rangle = 
(25.6\pm 2.9)$km/s for the sample $W^{14}_{12}$ up to 
$\langle b\rangle = (38.4\pm 2.7)$km/s for 660 absorbers  
with $N_{HI}\geq 10^{14}{\rm cm}^{-2}$ what indicates a weak 
correlation of observed properties of absorbers. Detailed 
discussion of observed  characteristics of absorbers can 
be found in Kim, Cristiani \& D'Odorico (2002), Kim et al. 
(2002). 

The small variations of $\langle b\rangle$ with the 
redshift are accompanied by similar small variations 
of dispersion $\sigma_b\sim 0.5\langle b\rangle$ which also 
do not exceed $\sim$ 8\%. This fact indicates that the 
broad distribution function of $b$ is weakly dependent upon 
the redshift. The same is valid for the depth of potential 
wells, $\langle \Delta\Phi\rangle$ (\ref{phi}), and 
illustrates a correlation between the DM column density, 
$q$, and the overdensity of absorbers, $\delta$ described 
by (\ref{main}). The nature of such a  weak redshift evolution 
of these statistical characteristics of absorbers remains 
a mystery. 

The function $\langle G_{12}\rangle/G_0$ is also 
plotted in Fig. 2 together with the rough fit (\ref{g12z}). 
Its weak redshift dependence is a natural consequence of the 
weak redshift variations of $\langle b\rangle$ and $\langle 
lg N_{HI}\rangle$. Its deviations from (\ref{g12z}) correlate 
with the absorbers distribution plotted in Fig 1\,.  

\subsection{Redshift variations of the mean DM column 
density, overdensity and entropy of absorbers}

Other characteristics of absorbers depend upon the physical 
model. For the model discussed 
in Sec. 2 (Eq. \ref{main}) with the function $\langle G_{12}
\rangle$ (\ref{g12}, \ref{g12z}) the redshift variations of 
the mean DM column density, $\langle\xi\rangle$, the mean 
entropy and overdensity, $\langle F_s\rangle$ and $\langle 
\delta\rangle$, are plotted in Fig. 3. 

The mean DM column density of absorbers, $\langle q\rangle\,
\&\,\langle\xi\rangle\approx \langle q\rangle(1+z)^2$, is 
the most stable parameter. In principle, $\langle\xi\rangle$ 
does not change due to the formation and merging of absorbers 
and due to their transverse compression and/or expansion (DD02). 
The differences between the expected and measured $\langle\xi 
\rangle$ characterize, in fact, the influence of disregarded 
factors, mainly the completeness and representativity of the 
samples, the pancake relaxation and the scatter of the function 
$G_{12}$. 

\begin{figure}
\centering
\epsfxsize=7.cm
\epsfbox{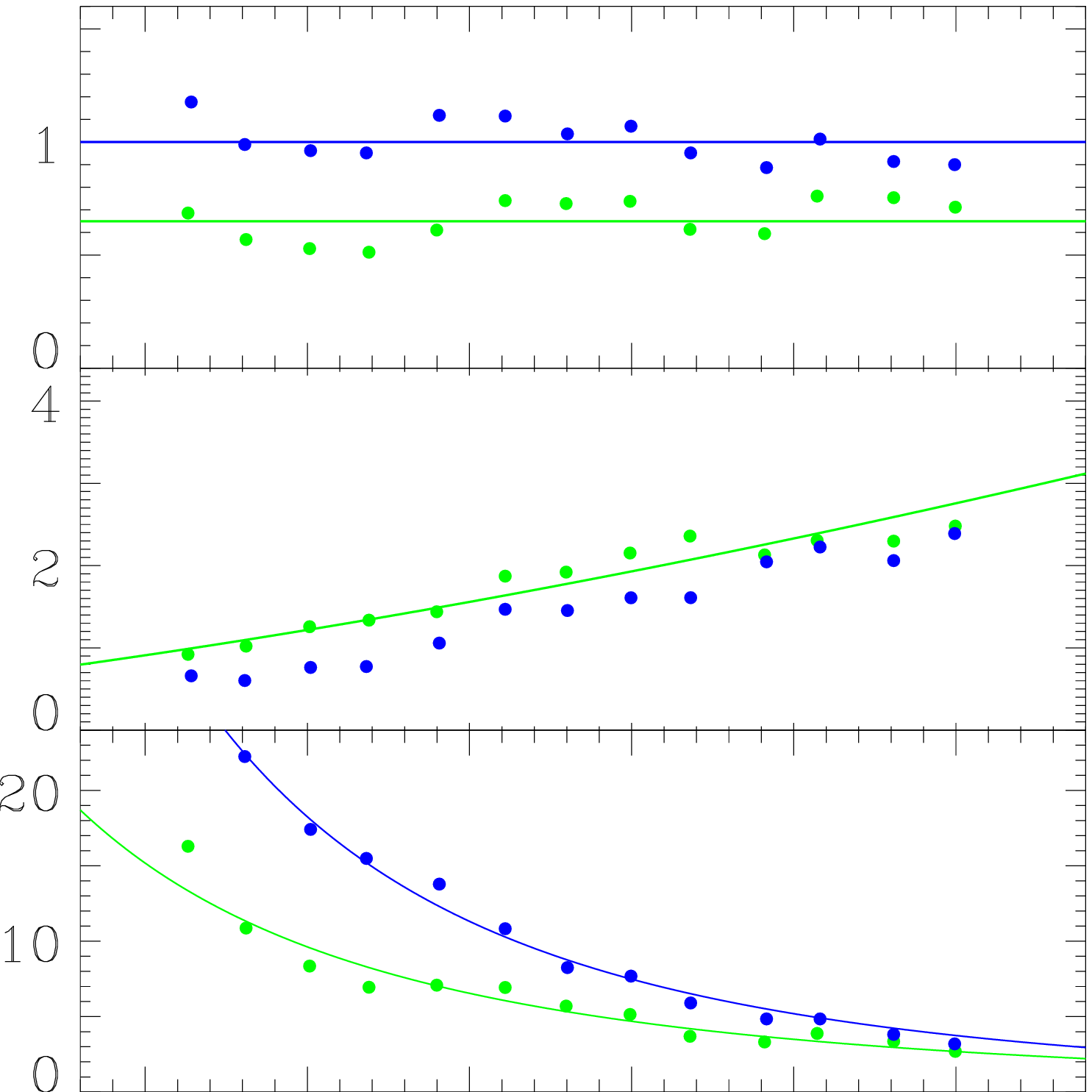}
\vspace{1.cm}
\caption{Functions $\langle\xi\rangle=(1+z)^2\langle q\rangle$ 
(top panel), $\langle F_s\rangle$ (middle panel) and the 
overdensity $\langle \delta\rangle$, (bottom panel) are plotted 
vs. redshift, $z$, for samples $S_{14}^{12}$ (green points), 
and $S_{18}^{13}$ (blue points). Fits (\ref{qttt}, \ref{linf}) are 
plotted by lines.
} 
\end{figure}

As is seen from comparison of Fig. 1 and Fig. 3, for both 
samples $S_{18}^{13}$ and $S_{14}^{12}$ variations of 
measured $\langle\xi(z)\rangle$ around the mean values  
\be
\langle\xi\rangle \approx (1.\pm 0.09),\quad 
\langle\xi\rangle \approx (0.65\pm 0.09)\,,
\label{qttt}
\ee
are roughly correlated with the redshift distribution of 
absorbers plotted in Fig. 1 what emphasizes the limited 
representativity of the sample used in the analysis. At small 
redshifts, $z\leq$ 2, the influence of the observational 
restrictions (Sec. 2.5) increase the measured value of 
$\langle\xi\rangle$.

For the sample $S_{14}^{12}$ the measured $\langle\xi(z)
\rangle\approx 0.65$ is close to the expected value 
$\langle\xi(z)\rangle\approx 0.82$ what verifies the choice 
of the amplitude of $\langle G_{12}\rangle$ in (\ref{g12z}). 
However, the estimates of $\langle\xi(z)\rangle$ depend 
also upon the choice of the amplitude of initial 
perturbations, $\tau_0$ (\ref{tau0}), known only with 
a moderate precision. For the samples $S_{18}^{13}$\,, 
$\langle\xi(z)\rangle$ increases due to the rejection 
of weaker absorbers in this sample. 

For both samples, $S_{14}^{12}\,\&\,S_{18}^{13}$, the minimal 
DM column density is found to be $q_{min}\sim  10^{-2}$  
and it only weakly depends upon the redshift. This result 
is especially sensitive to the random noise, it is based 
on poor statistic and is not reliable because our fit of 
$\langle G_{12}\rangle$ provides a reasonable description 
for the sample as a whole only. This value is smaller than 
the estimate $\langle q_{thr}\rangle\approx 3\cdot 10^{-2}$ 
(\ref{est}) obtained in Sec. 6.3 from the analysis of 
redshift distribution of absorbers. This disagreement 
requires more detailed investigation and, perhaps, 
redefinition of the factors $G_0$, $\Theta_H$ and 
$\Theta_\delta$ . 

As is seen from Fig. 3, the redshift variations of 
functions $\langle\delta\rangle$ and $\langle F_s\rangle$ 
characterizing the overdensity of DM component and entropy 
of gaseous component for both samples $S_{18}^{13}$ and 
$S_{14}^{12}$ can be roughly fitted as follows: 
\be
\langle\delta\rangle\approx 7.5 \zeta^{-3.1},\quad 
\langle\delta\rangle\approx 4.7 \zeta^{-2.5}\,,
\label{linf}
\ee
\[
\langle F_s\rangle\approx 1.93 \zeta^{1.6},\quad 
\zeta=(1+z)/4\,.
\]
These fits emphasize the general tendencies of absorbers 
evolution. Comparison with the background density (\ref{bg}) 
and entropy (\ref{sbg}) shows that the mean density of 
absorbers weakly depends on the redshift and the mean 
entropy weakly increases with time. These results 
indicate that our sample is dominated by long lived 
relaxed absorbers.  

These estimates may be rescaled according to (\ref{gam}). 

\subsection{The mean size of absorbers}

Our model of absorbers ( see Sec. 2) allows also to estimate 
roughly the real size 
of absorbers along the line of sight, $\Delta r$. Due to the 
strong correlation between $q$ and $\delta$, this size should 
be obtained by averaging $\Delta r$ as given by (\ref{delr}). 
For the model parameters given in (\ref{cosm}) and for both 
samples, $S_{14}^{12}$ and $S_{18}^{13}$, we have 
\be
\langle\Delta r\rangle\sim (80-100) h^{-1}{\rm kpc}\,,
\label{rdl}
\ee
and this size increases up to $\sim 150 h^{-1}$ kpc for weaker 
absorbers of the sample $W_{14}^{12}$. For all samples the PDFs 
of the sizes are similar to the Gaussian function with a dispersion 
$\sigma_r\sim 0.5\langle\Delta r\rangle$. 

The mean size of absorbers in the redshift space is associated with 
the Doppler parameter, $b$, as follows:
\[
\langle\Delta r_b\rangle= 2 \langle b\rangle/H(z)\approx 130
h^{-1}{\rm kpc}\left({4\over 1+z}\right)^{3/2}
\sqrt{0.3\over\Omega_m}\,.
\]
As usual, it is larger than that in the real space 
(\ref{rdl}). This difference agrees with the domination 
of relaxed gravitationally bound absorbers in the sample. 

The expected mean transverse size of absorbers is comparable 
with their Lagrangian size along the line of sight and was 
roughly estimated in DD02 as 
\be
\langle\Delta r_t\rangle\approx 5.2{l_v\tau^2\over 1+z}\approx 
{0.6 l_v\over (1+z)^{3}}\approx 300 h^{-1}{\rm kpc}\left({
4\over 1+z}\right)^3\,.
\label{rt}
\ee 
Difference between estimates $\langle\Delta r\rangle$ and 
$\langle\Delta r_t\rangle$ agrees with expectations of the 
Zel'dovich theory and indicates that such absorbers could 
be unstable with respect to the disruption into a system 
of denser less massive clouds with $\Delta r_t\sim \Delta 
r$ (see, e.g., Doroshkevich 1980; Vishniac 1983). 

\subsection{The mean matter fraction accumulated by 
absorbers and the mean absorber separation} 

The mean matter fraction accumulated by absorbers, 
$\langle f_{abs}(z)\rangle$, and the mean absorber 
separation, $\langle D_{sep}(z)\rangle$, defined by 
(\ref{sep}\,\&\,\ref{fract}), are plotted in Fig. 4 
for the samples $W_{14}^{12}$, $S_{14}^{12}$~\&~$S_{18}^{13}$. 
The estimates of $\langle f_{abs}(z)\rangle$ are 
sensitive to small measured line separations, $\Delta 
z\sim 10^{-4}$, which are unreliable. For richer absorbers 
of the sample $S_{18}^{13}$, the estimates of $f_{abs}(z)$ based
on the expression (\ref{fract}) are also unreliable because 
they neglect the transverse compression and expansion of 
matter and formation of high density clouds. So, 
the representativity and precision of measured $\langle 
f_{abs}(z)\rangle$ are limited. The estimates of $\langle 
D_{sep}(z)\rangle$ are less sensitive to the first factor 
but are also influenced by the second one. The random scatter 
of the measured $\langle f_{abs}(z)\rangle$ and $\langle 
D_{sep}(z)\rangle$ could be caused by the limited 
representativity of the samples at small and higher 
redshifts. 

\begin{figure}
\centering
\epsfxsize=7.cm
\epsfbox{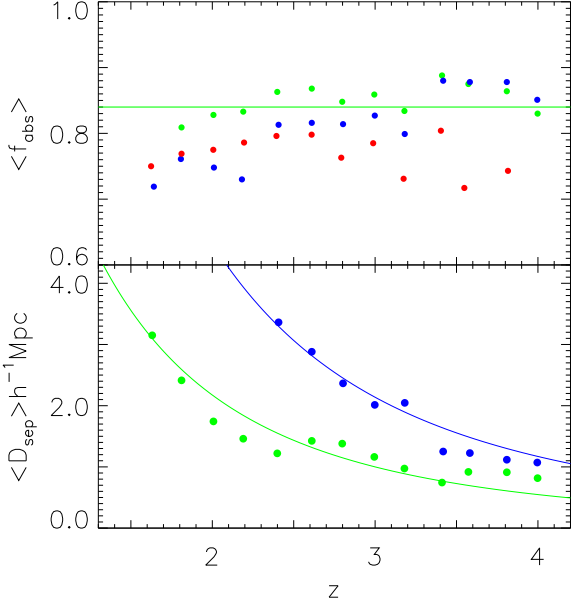}
\vspace{1.cm}
\caption{Mean fractions of matter accumulated 
by absorbers, $\langle f_{abs}\rangle$, (top panel) and 
the mean absorber separations, $\langle D_{sep}\rangle$,
(bottom panel), are plotted vs. redshift, $z$, for samples 
$W_{14}^{12}$ (red points), $S_{14}^{12}$ (green points), and 
$S_{18}^{13}$ (blue points). Fits (\ref{d1318}) are plotted by 
solid and dashed lines.
} 
\end{figure}

In spite of uncertainties these results are interesting 
in some respects. Thus, the regular growth of $\langle 
f_{abs}(z)\rangle$ and decrease of $\langle D_{sep}(z)
\rangle$ with redshifts for the sample $S_{18}^{13}$  
indicate the progressive disruption of richer absorbers 
and their transformation into system of high density clouds. 
Indeed, due to small cross--section of such clouds, they 
are rarely cross the line of sight and their formation 
increases the separation of richer observed absorber and 
even can decrease the measured $\langle f_{abs}\rangle$ 
for this sample. 

Both the significant fraction of matter seen as  numerous 
weaker absorbers in the sample $W_{14}^{12}$ and its 
weak redshift variation point in favor of the domination 
of absorbers merging as compared with the formation 
of new absorbers from the background matter. The merging 
remains unchanged the $\langle f_{abs}(z)\rangle$ but 
progressively increases $\langle D_{sep}(z)\rangle$
with time. These variations can be enhanced by the 
observational restrictions, $N_{HI}\geq N_{thr}$, 
discussed in Sec 2.5\,. 

For all samples the fraction $\langle f_{abs}(z)
\rangle\sim 0.8$ exceeds the theoretical expectation 
$\langle f_{abs}(z)\rangle\sim$~0.5--0.6 for $z\simeq 3$ 
(DD02 and Sec. 7.6) and is close to the maximal fraction, 
$f_{max}\simeq 0.875$, which can be compressed in the 
Zel'dovich' theory. This disagreement indicates the 
limited applicability of the one dimensional expression 
(\ref{fract}) and the limited precision of our model of 
absorbers. These results must be tested with richer samples 
of observed absorbers and by comparison with simulations. 
  
As is well known, at redshifts $z\geq$ 4 the absorbers 
begin to overlap and their separation becomes problematic. 
Analysis of the evolution of the mean absorbers separation 
allows to estimate quantitatively this effect. For this 
purpose, we can compare the mean separation of absorbers 
at various redshifts, $\langle D_{sep}(z)\rangle$, given 
by (\ref{sep}) with the mean Doppler parameter which 
characterizes the observed thickness of absorbers. For 
samples $S_{14}^{12}$ and $S_{18}^{13}$ 
these parameters are roughly fitted by power laws: 
\[
\langle D_{sep}\rangle\approx \left({4\over 1+z}\right
)^{2.7} h^{-1}{\rm Mpc},\quad \approx \left({5.3
\over 1+z}\right)^{2.7} h^{-1}{\rm Mpc}\,,
\]
\be
{H_0\langle D_{sep}\rangle\over\langle 
b\rangle}\approx \left({6.1\over 1+z}\right)^{2.6},\quad 
{H_0\langle D_{sep}\rangle\over\langle b\rangle}\approx 
\left({8\over 1+z}\right)^{2.9}\,,
\label{d1318}
\ee
respectively.

These results verify that at $z\geq$~4 absorbers  
effectively overlap and the system of individual absorbers 
is transformed into continuous absorption imitating the 
Gunn--Peterson effect. 

\subsection{Distribution functions of absorber parameters}

The observed probability distribution function, PDF, of the 
DM column density of absorbers can be compared with the 
expected one (\ref{qpdf}). However, we have no theoretical 
expression for the PDFs of the Doppler parameter, 
overdensity and entropy of absorbers because the 
action of random factors cannot be satisfactory 
described. 

Indeed, the relaxation of DM pancakes depends upon unknown 
internal structure of pancakes, the adiabatic compression 
and/or expansion of an absorber changes its overdensity and 
temperature. On the other hand, the radiative cooling and 
bulk heating lead to the drift of the gas entropy and 
overdensity but leaves unchanged the depth of potential 
well formed by DM distribution. Merging of pancakes 
increases more strongly the depth of potential well 
and the gas entropy but the overdensity of the gaseous 
component increases only moderately. All the time, the 
temperature and overdensity of trapped gas are rearranged 
in accordance with the condition of hydrodynamic equilibrium 
across the pancake. 

Because of this, our discussion of absorbers evolution 
has only phenomenological character. The PDF of the DM 
column density, $N_q$, is plotted in Fig. 5 for the sample 
$S_{14}^{12}$ together with the fits (\ref{qpdf}) 
and (\ref{q0pdf}). The distribution functions of the 
reduced Doppler parameter, $\upsilon$, are plotted in Fig. 
5 for samples $S_{18}^{13}$ and $S_{14}^{12}$ together with 
the Gaussian fits. 

\begin{figure}
\centering
\epsfxsize=7.cm
\epsfbox{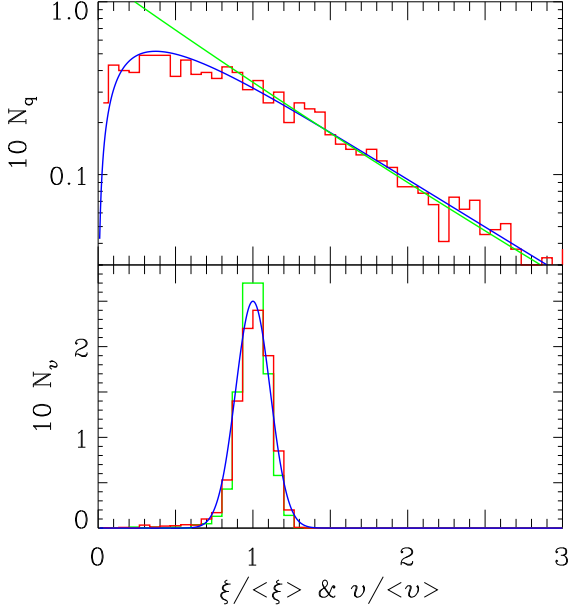}
\vspace{1.cm}
\caption{Top panel: PDFs of the DM column density, 
$N_q$, for the sample $S_{14}^{12}$. Fits (\ref{qpdf}) 
and (\ref{q0pdf}) are plotted by green and blue lines. 
Bottom panel: PDF $N_\upsilon$, for the samples $S_{14}^{12}$ 
(red line) and $S_{18}^{13}$ (green line). Gaussian fit is 
plotted by blue line. 
} 
\end{figure}

\subsubsection{Distribution functions of the DM column  
density}

The PDF for the DM column density of pancakes, $N_q(q)$, 
plotted in Fig. 5 for samples $S_{14}^{12}$ is the most 
interesting because it can be compared with the theoretically 
expected PDF (\ref{qpdf}) which, in particular, is sensitive 
to the coherent length of initial density field, $q_0$. As 
is seen from (\ref{qpdf}), the PDF $N_q(\xi)$, with 
$\xi\approx q(1+z)^2$, does not depend on the redshift 
and, so, the joint PDF can be obtained for absorbers observed 
at all redshifts. 

For $\xi\geq 0.5\langle\xi\rangle$, the PDF plotted in Fig. 
5 is well fitted by the function (\ref{qpdf}) for $q_0=0$ 
and $\langle\xi\rangle =$ 0.57 what is close to $\langle\xi
\rangle =$ 0.65 (\ref{qttt}). The deficit of observed 
absorbers with $\xi\leq\langle\xi\rangle$ can be related to 
incompleteness of the observed sample for small $\xi$. This 
result verifies the self consistency of the physical model 
used here and the Gaussianity of initial perturbations. 

However, as it is seen from (\ref{qpdf}), this deficit of 
weaker absorbers can be related also to the suppression of 
the formation of poorer pancakes caused by the correlations 
of small scale initial perturbations and Jeans damping 
described by the parameter $q_0$ in (\ref{qpdf}). 
Indeed, the observed PDF is well fitted also by the function 
\be
N_q= 0.18~{{\rm erf}(\sqrt{y})\over\sqrt{y}}
{\exp(-y)\over (1+0.53/y)},
\quad y=1.15{\xi\over\langle\xi\rangle}\,,
\label{q0pdf}
\ee
what is consistent with rough estimate $q_0\sim$~0.05. This 
value depends on the statistics of absorbers at small 
$\xi\leq\langle\xi\rangle$ and can be considered as an upper 
limit of $q_0$ only. 

\subsubsection{Distribution functions of the reduced 
Doppler parameter}

As was noted above, the theoretical description for the PDFs 
of the reduced Doppler parameter, overdensity and entropy of 
the gas is problematic because they depend upon many random 
factors. Therefore, here we will restrict our analysis to the 
fits of observed PDFs, $N_\upsilon$, taking also into account 
the correlations of the measured $\xi,\,b,\,\delta\,\&\,F_s$. 

To adjust the reduced characteristics of pancakes introduced 
in Eq. (\ref{reduct}), we will minimize the correlation 
of $\xi$ with $\upsilon,\,\Delta,\,\&\,S$. For the sample 
$S_{14}^{12}$, the adjusted reduced characteristics are defined 
as follows:
\[
\gamma=1.8,\quad \upsilon = \ln(\beta\xi^{-0.44})\,,
\]
\be
\Delta = \ln[(1+z)^3\delta~\xi^{-1.1}] = \ln\delta_0-2 
\upsilon\,,
\label{ered}
\ee
\[
S=\ln[(1+z)^{-2}F_s\xi^{-0.15}] = const. +3.33~\upsilon\,.
\]
These expressions
correspond to Eq. (\ref{reduct}) with the effective power 
index of compressed DM component $\gamma=1.8$. For the 
sample $S_{18}^{13}$ we have $\gamma=2.1$. Both values 
are close to $\gamma=2$ discussed in Sec. 2.3.1\,. With 
such definition, we get the negligible correlations, $\leq
0.01$, between $\xi$ and other reduced characteristics and 
a strong correlation $\approx$ 1 between $\upsilon$ and 
$\Delta\,\&\,S$. The distribution function $N_\upsilon$ 
plotted in Fig. 5 is well fitted by the Gauss function 
with $\langle\upsilon\rangle=0.81\pm 0.04,~\sigma_\upsilon
\approx 0.11\langle\upsilon\rangle$\, what is consistent 
with the assumption of approximate equilibrium of majority 
of the observed absorbers. 

These results mean that, in fact, the joint action of all 
random factors is characterized by the one random function 
$\upsilon$ which can be directly expressed through the 
observed parameters as follows
\be
\upsilon\approx 0.48 \ln~\beta-0.15\ln(N_{HI}/N_0/\delta_0),
\quad \gamma=1.8\,.
\label{onbh}
\ee 
More detailed analysis requires the discrimination of  
subpopulations of absorbers with different evolutionary 
histories.  

\subsection{Three subpopulations of absorbers}

\begin{table}
\caption{Parameters of three subpopulations of absorbers for 
the 
sample $S_{14}^{12}$}
\label{tbl2}
\begin{tabular}{lrc ccc cr } 
   &$N_{abs}$&$\langle f_s\rangle$&$\langle lg N_{HI}
\rangle$&$\langle b\rangle$&$r_{bH}$&$\gamma$&$\langle
\upsilon\rangle$\cr
\hline
   full   &4369&1.00&13.1&29&0.30&1.8&0.8\cr 
   sh     &1940&0.44&13.0&37&0.33&1.8&1.0\cr  
   cl     &1664&0.38&13.5&23&0.67&1.6&0.6\cr  
   exp    & 765&0.18&12.3&18&0.42&3.5&1.3\cr  
\hline
\end{tabular}

$f_s$ is the fraction of absorbers in a subpopulation,
$r_{bH}$ is the linear correlation coefficient of $b$ and 
log$N_{HI}$. 
\end{table}

As a first step of more detailed investigation of 
absorbers evolution, we can approximately separate three 
subpopulations of absorbers with different overdensities 
and entropies. Due to continuous distribution of both 
$\delta\,\&\,F_s$ for the full samples and because both these
values depend upon unknown random factors (\ref{gam}), 
the separation is quite arbitrary and characteristics 
of subpopulations depend upon the sample used in the analysis and 
the parameters (\ref{cosm}). Even so, the redshift 
variation of mean parameters of subpopulations allows 
to trace roughly their evolutionary history and to 
describe the main stages of absorbers formation and evolution. 

The first subpopulation, let us call it "sh", with 
$\delta\geq$~1 and $F_s\geq$~1 is formed due to the 
merging and shock compression. The second one, called 
here "cl", includes low entropy absorbers with $\delta
\geq$~1, $F_s\leq$~1\,. The third one, called here "exp", 
contains peculiar absorbers with $\delta\leq$~1. 

This choice of the threshold parameters allows to separate 
absorbers formed and observed at the same redshifts, 
$z_f\approx z_{obs}$. However, due to evolution of 
the background density (\ref{bg}) and entropy (\ref{sbg}), 
the functions $\delta\,\&\,F_s$ defined by (\ref{main}) are 
shifted with time even when the overdensity and entropy 
of a long lived absorber are not changed. This shift leads 
to some artificial intermixture of subpopulations. 

For the sample $S^{12}_{14}$ the mean parameters of these 
subpopulations are listed in Table 2, where $r_{bH}$ is 
the linear correlation coefficient of $b$ and $\lg N_{HI}$. 
Redshift variations of $\langle b\rangle$, $\langle 
lgN_{HI}\rangle$, $\langle F_s\rangle$, $\langle\delta
\rangle$, $\langle q\rangle$, $\langle \xi\rangle$ and 
the relative number fractions of absorbers in these 
subpopulations, $\langle f_m\rangle$, are plotted in 
Figs. 6 \& 7 and given in (\ref{sh}, \ref{cl}\,\&\,\ref{exp}). 

\begin{figure}
\centering
\epsfxsize=7.cm
\epsfbox{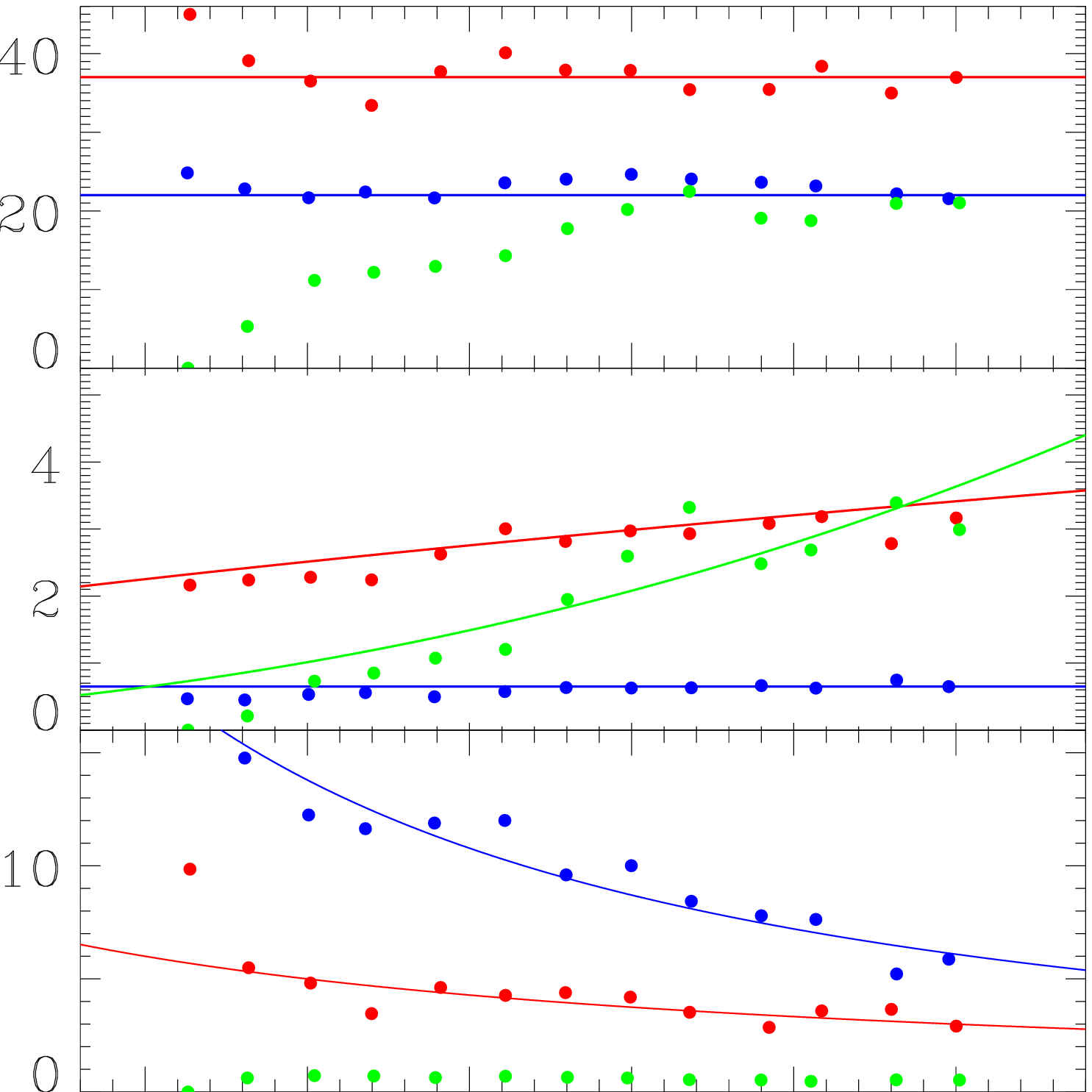}
\vspace{1.cm}
\caption{The mean Doppler parameter, $\langle b\rangle$ 
(top panel), the mean entropy, $\langle F_s\rangle$ (middle 
panel), and the mean overdensity $\langle\delta\rangle$ 
(bottom panel), are plotted vs. redshift, $z$, together 
with fits (\ref{sh}, \ref{cl},\,\&\,\ref{exp}) 
for the "sh" (red), "cl" (blue) and "exp" (green) 
subpopulations of the sample $S_{14}^{12}$\,.
} 
\end{figure}

The subpopulation "sh" is composed of the richest and 
hottest absorbers with
\be
\langle lg N_{HI}\rangle\approx 13.0\pm 0.4,\quad \langle 
b\rangle\approx (37\pm 2.7)km/s\,,
\label{sh}
\ee
\[
\langle\xi\rangle\approx 0.8\pm 0.15,\quad 
\langle F_s\rangle\approx 3.0\zeta^{0.6},
\quad \langle\delta\rangle\approx 3.75/\zeta\,,
\]
where $\zeta=(1+z)/4$. As is seen from Fig. 7, for this 
subpopulation the decrease of the relative number fraction of 
absorbers, $\langle f_m\rangle$, with $z$ is accompanied 
by the growth of the DM column density, $\langle q\rangle$. 
Such variations are typical for absorbers formed due to 
pancake merging and strong shock compression of the matter. 

The subpopulation "cl" includes absorbers with 
\be
\langle lg N_{HI}\rangle\approx 13.5\pm 0.37,\quad 
\langle b\rangle\approx (23.1\pm 1.1) {\rm km}/{\rm s}\,,
\label{cl}
\ee
\[
\langle\xi\rangle\approx 0.69\pm 0.1,\quad \langle F_s
\rangle \approx 0.6\pm 0.08,\quad \langle\delta\rangle
\approx 8.7 \zeta^{-1.6}\,.
\]
Such behavior of the mean characteristics indicates that 
this subpopulation can be dominated by absorbers formed 
in low entropy regions at $z_f\sim z_{obs}$ due to adiabatic 
and weak shock compression. Such compression provides higher 
overdensity than the strong shock compression and does not 
change the entropy of compressed matter. For these absorbers 
the radiative cooling can also be more important. This 
subpopulation includes also some fraction of long lived 
absorbers formed at $z_f >z_{obs}$ due to merging and 
strong shock compression. This intermixture is caused by 
redshift variations of background parameters (\ref{bg}) 
and (\ref{sbg}). 

The peculiar subpopulation "exp" accumulates poorer 
absorbers with 
\be
\langle lg N_{HI}\rangle\approx 12.3\pm 0.3,\quad 
\langle b\rangle\approx (18\pm 5.4){\rm km}/{\rm s},
\label{exp}
\ee
\[
\langle q\rangle\approx 1.1\cdot 10^{-2},\quad  
\langle F_s\rangle\approx 2.1\zeta^{2.5},
\quad \langle \delta\rangle\approx 0.6\pm 0.08\,.
\]
Such behavior of $\langle q\rangle,\,\langle\delta\rangle\,
\&\,\langle F_s\rangle$ shows that these absorbers could be 
related to the unstable poorer pancakes formed within low 
density regions (see, e.g., Zhang et al. 1998). They are 
observed mainly at higher redshifts, they expand together with 
the background and disappear at low redshifts. Large values 
of the factor $\gamma=$~3.5 and of the reduced Doppler 
parameter $\langle\upsilon\rangle=$~1.3 (Table 2), verify 
the peculiar character of this subpopulation. 

\begin{figure}
\centering
\epsfxsize=7.cm
\epsfbox{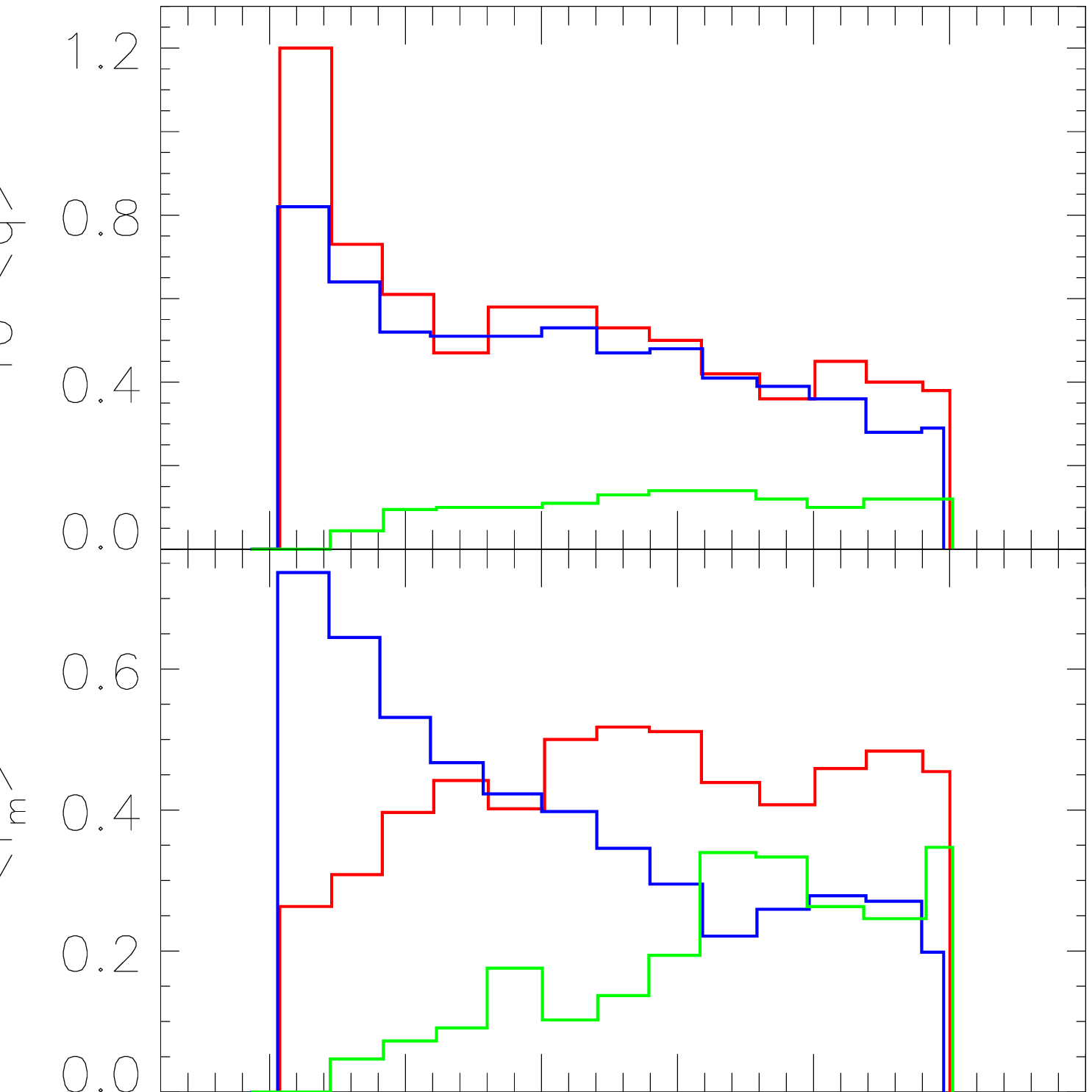}
\vspace{1.cm}
\caption{The DM column density (top panel) and the relative 
number fractions of absorbers (bottom panel) for the "sh" 
(red lines), "cl" (blue lines) and "exp" (green lines) 
subpopulations  of the sample $S_{14}^{12}$ are plotted 
vs. redshift, $z$ . 
} 
\end{figure}

Moreover, the analysis performed in Sec. 6.3 indicates 
that, for weaker absorbers dominating this subpopulation, 
at least the DM column density, $\langle q\rangle$, can 
be underestimated. This means that other parameters of 
the subpopulation can also be estimated unreliably. In 
particular, if these absorbers are unstable then the 
random factors become time dependent and the estimates 
(\ref{exp}) must be corrected. Fortunately, the fraction 
of such absorbers is small and its influence on other 
estimates is limited.

In spite of the quite arbitrary separation of subpopulations 
these results illustrate the action of evolutionary factors 
mentioned above. The objective character of the discrimination 
is seen from the comparison of observed parameters, $\langle 
lg N_{HI}\rangle$ and $\langle b\rangle$, and their linear 
correlation coefficient, $r_{bH}$, listed in Table 2 for 
the full samples and for the separated subpopulations. 

These results confirm that majority of observed absorbers 
represent gravitationally bounded pancakes formed in the 
course of adiabatic and shock compression at $z_f\geq 
z_{obs}$. They demonstrate similarity of evolutionary 
history and generic origin of subpopulations "sh" and 
"cl". More detailed statistical investigation of 
absorbers evolution requires finer discrimination of 
subpopulations and therefore richer set of observed 
absorbers is needed.

\subsection{Absorbers as a test of background properties} 

Some absorbers characteristics can be used to estimate 
the redshift variations of mean properties of homogeneously 
distributed hydrogen (see, e.g., Hui \& Gnedin 1997; Schaye 
et al. 1999, 2000; McDonald et al. 2001). Thus, weak redshift 
variations of $\langle F_s\rangle$ (\ref{cl}) suggest also
weak redshift variations of $T_{bg}$ (\ref{bg}). 
However, such estimates  are inevitably approximate and their 
significant scatter is caused by action of many factors 
discussed above. 

To obtain more stable results Schaye et al. (1999, 2000) 
consider a cutoff at small $b$ in the distribution of
$b(N_{HI})$. Indeed, relations (\ref{main}) can be 
rewritten as follows:
\be
b=b_{bg}\kappa_b^{3/8}\delta_0^{1/8}\left({F_s\over (1+z)^2}
\right)^{9/16}\left({N_{HI}\over N_0}\right)^{1/4}\,.
\label{adb}
\ee
Parameters $N_0, \delta_0, b_{bg}\,\&\,\kappa_b$ were 
defined by (\ref{bg}, \ref{bb},\,\&\,\ref{NH2}).
Evidently, for $F_s={\rm const.}$ this relation is identical 
to the equation of state $b^2={\rm const.}\delta^{2/3}$. It 
demonstrates that, for the subpopulation of absorbers 
formed at $z_f\approx z_{obs}$ with the same $F_s$, 
the observed parameters $z, b$ and $N_{HI}$ are strongly 
correlated and, for a given $N_{HI}$, $b$ decreases with 
$F_s$. However, statistics of absorbers near the cutoff is 
small, the action of factors discussed above erodes the 
cutoff and makes its interpretation less reliable.  

For absorbers selected from the sample $S^{12}_{14}$ in 
four redshift intervals, $z\leq$~2~$\leq z\leq$~2.5~$\leq 
z\leq$~3~$\leq z$, the boundary of the observed distribution 
$b(N_{HI})$ can be roughly fitted by the relation 
\be
b\approx b_0(1+z)^{-1}(N_{HI}/10^{12}cm^{-2})^{\gamma_b}\,.
\label{bfit}
\ee
For $z\leq$~3 we have $\gamma_b\sim$ 0.25 that coincides 
with (\ref{adb}) but the value of $b_0\sim$~22 km/s is 
$\sim$~2 times smaller, what is expected from (\ref{adb}). 
This means that in the framework of the model under 
consideration the cutoff can be naturally related to a 
fraction of low entropy absorbers with $F_s\sim$~0.3 rather 
then with $F_s\sim$~1. These absorbers can be formed at 
$z_f\sim z_{obs}$ due to the adiabatic compression within 
low entropy regions or due to both adiabatic and shock 
compressions at redshifts $z_f\geq~1.5~z_{obs}$. For richer 
absorbers with $lg N_{HI}\geq$~13 the radiative cooling is 
also more important. This explanation is quite consistent 
with the estimates (\ref{cl}).

At $z\geq$~3 we have for the fit parameters $\gamma_b
\sim$~0.15--0.17, $b_0\sim$~38km/s, what differs from  
(\ref{adb}). This difference can be partly related to the 
poor statistics at higher redshifts and 
to  admixture of long lived absorbers formed at redshifts 
$z\geq$~4 when stronger systematic and random variations 
of UV background are expected. In particular, at $z\sim$~3 
$He II$ becomes ionized and the spectrum of UV background 
and the effective power--low index of the 
temperature--density relation are changed (Songaila 1998; 
Schaye et al. 2000; Theuns et al. 2002a, b). 

These results demonstrate that, in the framework of 
model under consideration, the cutoff in the distribution 
of $b(N_{HI})$ can be explained  but its interpretation is 
not so clear. The problem requires further investigation.  

\section{Walls at high redshifts}

Majority of the rare absorbers with higher Doppler parameter, 
$b\geq$ (80 -- 90)km/s and moderate column density of neutral 
hydrogen $N_{HI}\leq 10^{14.5}cm^{-2}$, can be identified with 
the embryos of richer structure elements which are seen now 
as rich walls in the observed galaxy distribution. The number 
of such absorbers is small and their statistical characteristics 
cannot be reliably determined. However, for 45 such absorbers 
with $b\geq$ 90km/s selected in 9 spectra at $1.8\leq z\leq 
3.8$ we have  
\be
\langle lg N_{HI}\rangle\sim 13.2,\quad \langle\xi\rangle\sim 
1.45,\quad \langle\delta\rangle\sim 6,\quad \langle F_s\rangle
\sim 10\,,
\label{ww}
\ee
and the mean comoving separation of such absorbers (\ref{sep}) 
is  
\be
\langle D_{sep}\rangle\approx (60\pm 20)\sqrt{0.3/
\Omega_m}h^{-1}{\rm Mpc}\,.
\label{dsep}
\ee

Such large separation suggests that these objects will retain 
their individuality up to small redshifts and, indeed, this 
separation is quite consistent with the mean wall separation 
$\langle D_w\rangle\approx (66\pm 13) h^{-1}$ Mpc measured for 
the SDSS EDR at $z=0$ (Doroshkevich, Tucker \& Allam 2002). 
However, for smaller $b$, the number of absorbers rapidly 
increases and already for 139 absorbers with $b\geq$~70km/s 
in 15 spectra the mean separation decreases to $D_{sep}
\sim (40\pm 20)h^{-1}$Mpc. 

These results show that the walls begin to form already at 
$z\sim 3$ and continue to form up to the present time. At the 
same time, the continuous distribution of all parameters of 
absorbers shows that at high redshifts the discrimination of 
walls and other structure elements is quite arbitrary. The 
problem deserves further investigation in a wider range of 
redshifts with a more representative sample of absorbers 
especially at high redshifts. Perhaps, such walls can be 
also observed in the galaxy distribution at high redshifts. 

\section{Redshift distribution of absorbers}

In this section we compare the observed redshift 
distribution of absorbers with theoretical relations 
(\ref{nw},\,\ref{nq0}) describing the expected redshift 
evolution of the 1D number density of DM pancakes, $n_{
abs}$, due to their formation and merging. As is seen 
from these relations, the number density depends upon 
the cosmological model, moments of initial power spectrum 
and the threshold DM column density, $q_{thr}$. 
Therefore, to compare these expectations with observed 
absorbers distribution we have to select samples with 
$q\geq q_{thr}$ rather than to use samples with a minimal 
$N_{HI}$ discussed in Sec. 4. Such samples can be prepared 
using estimates of $q$ as given by (\ref{main}) in spite of 
their unreliability for smaller $q$ (see Sec. 4.2, 6.3). 

Both relations (\ref{nw}\,\&\,\ref{nq0}) are derived for 
the Gaussian initial perturbations. They depend on the 
redshift of absorbers only and, so, the impact of absorbers 
velocities and Doppler parameters on the resulting estimates 
is minimal. At $z\leq$~4 these factors increase the scatter 
of measured $n_{abs}$ but only at $z\geq$~4 -- 4.5 they lead 
to a significant overlapping of absorbers (Sec. 4.4). The 
Jeans damping caused by the temperature of background is 
well described by (\ref{nq0}) through the parameter $q_0$. 

Four samples, namely, $Q_{14}^{50}$, $Q_{14}^{20}$, 
$Q_{14}^{10}$, and $Q_{14}^{01}$, with $q_{thr}=$ 0.05, 
0.02, 0.01 \& 0.001 were selected from 14 high resolution 
spectra. These samples contain 1554, 3299, 3998 and 4475 
absorbers, respectively. Comparison of $q_{thr}$ used for 
the sample preparation with estimates of $\langle q_{thr}
\rangle$ obtained from fits (\ref{nw},\,\ref{nq0}) allows 
to test the self consistency of our physical model and 
the choice of parameters (\ref{cosm}). 

The richest sample $Q_{14}^{01}$ includes almost  
all lines and probably is incomplete. It can be used for 
comparison with results obtained for other samples. The 
sample $Q_{14}^{50}$ contains richer absorbers and is 
weakly sensitive to properties of small scale density 
field described by the parameter $q_0$ in (\ref{nq0}). 
It is used to test the expression (\ref{nw}) and to 
estimate $\langle q_{thr}\rangle$ for such absorbers. 
The samples $Q_{14}^{20}$, $Q_{14}^{10}$ can be used for 
independent estimates of both $\langle q_{thr}\rangle$ 
and $q_0$. Of course, these estimates are statistical 
and averaged over the sample. 

To test the sample dependence of our results we use the 
sample $Q_{18}^{10}$ with $q_{thr}=$ 0.01 selected from all 
18 spectra. As compared with the sample $Q_{14}^{10}$ it 
contains $\sim$10\% additional absorbers with $N_{HI}\geq 
10^{13} cm^{-2}$. To demonstrate the possible variations of 
intensity of UV background we consider also the sample 
$W_{14}^{12}$ containing 2126 weaker absorbers with 
$N_{HI}\leq 10^{12} cm^{-2}$. Here we 
also use the sample of HST data (Bahcall et al. 1993, 1996; 
Jannuzi et al. 1998) which contains 1000 absorbers at 
$z\leq$ 1.5 . This sample is probably incomplete at least 
at $z\geq$~1 . 

\subsection{Selection of Poissonian subsamples of absorbers}

The first problem is the selection of Poissonian subsamples 
of absorbers and estimation of their mean number density 
at various redshifts. For this purpose, we use the measured 
redshift separations of neighboring absorbers, $\Delta z$, 
what partly attenuates the influence of selection effects 
inherent in individual spectra. Instead of separation we 
use the dimensionless comoving distance,
\[
\Delta l = H_0\Delta z/H(z)\,,
\]
and absorbers in the interval $z-dz< z < z+dz$, taken 
from the sample under investigation, are organized into 
an `equivalent single field' by arranging the separations 
$\Delta l$ one after the other along the line of sight. 
For richer samples, distribution of absorbers separations 
obtained in this way is similar to the Poissonian one and 
the mean number density, $\langle n\rangle$, can be found 
by comparing the measured PDF with the differential or 
cumulative Poissonian PDFs: 
\be
dN/dx=\exp(-\langle n\rangle ~x)/\langle n\rangle\,,
\label{df}
\ee
\be
N=\exp(-\langle n\rangle ~x)\,.
\label{cf}
\ee

To decrease uncertainties and to estimate the actual scatter 
of these fits we rejected all points before the maxima of 
differential PDF and only more representative middle part 
of PDFs with the fraction of points $f\leq f_{thr}$ were 
used. The threshold fraction, $f_{thr}$, was varied between 
$f_{thr}=0.7$ and $f_{thr}=0.95$. Mean values and dispersions 
of the set of measurements with different $f_{thr}$ were 
taken as the actual value and scatter of $n_{abs}$. As a 
rule, the fit (\ref{cf}) of cumulative PDFs gives more 
stable estimates and smaller error bars than the fit 
(\ref{df}) of differential PDFs. Both estimates of $n_{abs}$ 
are plotted in Figs. 8\,\&\,9. 

\subsection{Variations of intensity of the UV background}

\begin{figure}
\centering
\epsfxsize=7.cm
\epsfbox{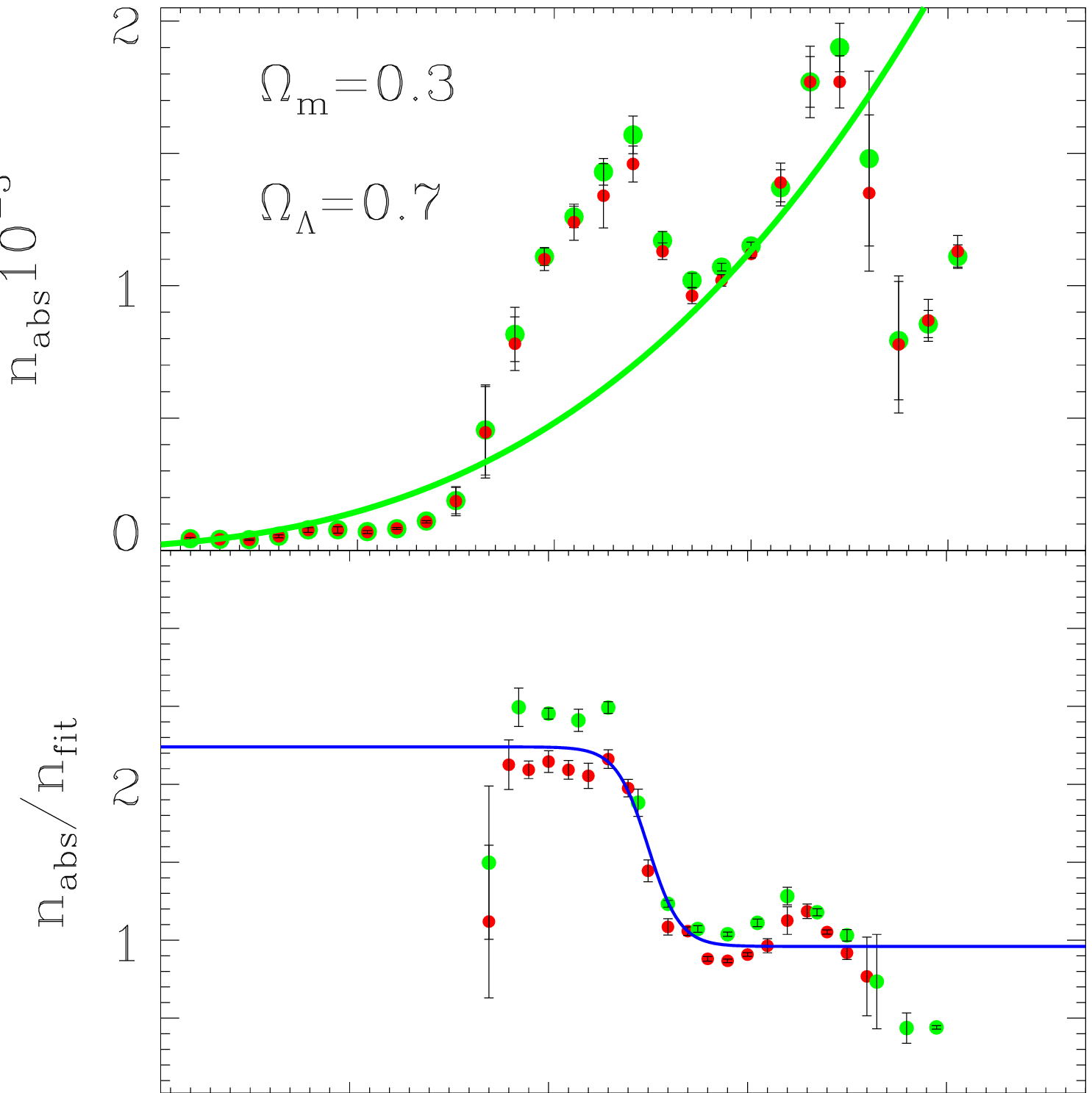}
\vspace{1.cm}
\caption{Top panel: redshift distribution of absorbers, 
$n_{abs}(z)$, found with fits (\ref{df}) and (\ref{cf}) 
(red and green points) for the sample $W_{14}^{12}$. Fit 
(\ref{nq0}) is plotted by green line.
Bottom panel: redshift variations of $n_{abs}/n_{fit}$ for 
the same sample. Fit (\ref{fgam}) is plotted by blue line.
} 
\end{figure}

For the sample $W_{14}^{12}$ the redshift distribution 
of absorbers is plotted in Fig. 8 together with the fit 
(\ref{nq0}) which characterizes the expected redshift 
distribution of absorbers. The weak absorbers in this 
sample are especially sensitive to variations of intensity 
of the UV background radiation what is manifested by the 
strong variations of $n_{abs}$ in comparison with the 
smooth fitting curve. This effect can be enhanced by the 
incompleteness of the sample and the nonhomogeneity of 
redshift distribution of observed weaker absorbers. At 
$z\leq$ 2 the rapid decrease of $n_{abs}$ is caused by 
the influence of the threshold of the observed column density 
of neutral hydrogen, $N_{HI}\geq N_{thr}\approx 10^{12}
cm^{-2}$, discussed in Sec. 2.5.

The observed variations of $\langle n_{abs}\rangle$ are 
well fitted by the function 
\be
\langle n_{abs}\rangle/n_{fit} = 1.6+0.65~{\rm th}\left
({2.5-z\over 0.16}\right)\,,
\label{fgam}
\ee
also plotted in Fig. 8. Similar variations with smaller 
amplitude are also seen for absorbers in other samples. 

The column density of neutral hydrogen is proportional to 
$\Gamma_\gamma^{-1}$ and an increase of $\Gamma_\gamma$ shifts 
weak absorbers below the observational threshold. This means 
that the relation (\ref{fgam}) can describe quite well the 
redshift dependence of variations of the UV background but their 
amplitude depends upon the distribution function of $N_{HI}$. 
Direct estimates show that the variations of the intensity 
by about 2 -- 3 times described by (\ref{zgamma}) are consistent with 
the observed variation of $\langle n_{abs}\rangle$. 
As was noted in Sec. 2.6 the influence of these variations 
on the mean characteristics of absorbers can be partly 
compensated by variations of random factors $\Theta_H\,\&\,
\Theta_\delta$.  

Variations of the UV background at $z\sim 3$ were already 
detected by different methods (see, e.g., Songaila 1998; 
Scott et al. 2000; McDonald\,\&\,Miralda--Escude 2001). 
This approach seems to be sufficiently perspective but 
quantitative results are sample dependent and must be 
tested with more representative and homogeneous samples 
of weak absorbers.  

\subsection{Redshift distribution of DM pancakes}

In this Section we consider the redshift distribution of 
DM pancakes for the samples $Q_{14}^{50}$, $Q_{14}^{20}$, 
$Q_{14}^{10}$, and $Q_{14}^{01}$ prepared with $q_{thr}(z)
\approx {\rm const.}$ for parameters (\ref{cosm}). For the sample 
$Q_{14}^{10}$, the function $n_{abs}(z)$ is plotted in Fig. 9 
together with the best fit (\ref{nq0}) and the random scatter 
of observed points around the fit. At redshifts 1.8~$\leq 
z\leq$~3.3 where the representativity of samples is better 
the scatter does not exceed $\sim$ 20\% . 

At redshifts $z\sim$ 1.5 -- 2 the rapid growth of observed 
mean number density of absorbers
\be
\langle n_{abs}\rangle\propto \left({1+z\over 2.7}\right)^8\,,
\label{n8}
\ee
coincides with the expected one (\ref{poor}) and is 
naturally explained by the influence of observational 
threshold $N_{HI}\geq N_{thr}\approx 10^{12}cm^{-2}$. 
Such cutoff is not seen for the subpopulation of richer 
absorbers with $N_{HI}\geq 10^{14}cm^{-2}$  what verifies 
its connection with the observational threshold. The probable 
incompleteness of the sample of weak absorbers at $z\leq$ 1.5 
enhances this rapid growth of $\langle n_{abs}\rangle$.
  
\begin{figure}
\centering
\epsfxsize=7.cm
\epsfbox{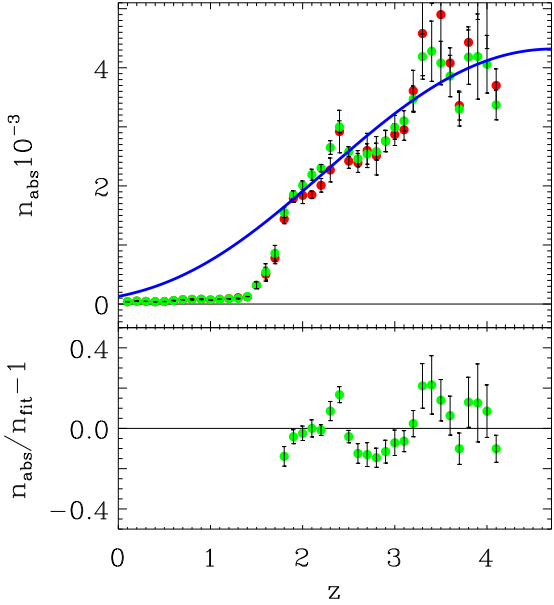}
\vspace{0.9cm}
\caption{Top panel: redshift distribution of absorbers, 
$n_{abs}(z)$, found with fits (\ref{df}) and (\ref{cf}) 
(red and green points) for the sample $Q_{14}^{10}$. Fit 
(\ref{nq0}) is plotted by the blue line.
Bottom panel: Differences between measured and fitted 
absorber distributions.  
} 
\end{figure}

The best estimates of the fit parameters (\ref{nq0}) for the 
samples $Q_{14}^{20}$, $Q_{14}^{10}$, and $Q_{14}^{01}$ are, 
respectively,
\[
\langle q_{thr}\rangle\approx (3.3\pm 0.3)\cdot 10^{-2},\quad 
\langle q_{thr}\rangle\approx (3.1\pm 0.3)\cdot 10^{-2}\,, 
\]
\be
\langle q_{thr}\rangle\approx (3.1\pm 0.3)\cdot 10^{-2}\,,
\label{est}
\ee
and for all three samples 
\be
q_0\approx (0.9\pm 0.1)\cdot 10^{-2}\,.
\label{q0est}
\ee
Here the formal errors of the fits are given. 

To test the sample dependence of the results we analyzed in the 
same manner the sample $S_{18}^{10}$ selected with $q_{thr}$=
0.01 from all 18 spectra (4404 absorbers). In this case, the 
excess of absorbers with $N_{HI}\geq 10^{13} cm^{-2}$ and 
larger separations decrease the measured $n_{abs}$ and 
the resulting estimates are 
\[
\langle q_{thr}\rangle\approx (3.2\pm 0.3)\times 10^{-2},
\quad q_0\approx (1.1\pm 0.12)\times 10^{-2}\,.
\]
We see that such perturbations lead to moderate variations 
of both $\langle q_{thr}\rangle$ and $q_0$\,.

Real precision of both estimates (\ref{est}) and 
(\ref{q0est}) depends upon several factors,  mostly on  
the limited representativity of the sample and precision 
of estimates $\tau_0$ (\ref{tau0}). As is seen from 
(\ref{nq0}), we really evaluate the product $q_{thr}
\tau_0^{-2}$ and $q_0\tau_0^2$, and, so, the real precision 
of both estimates (\ref{est}) and (\ref{q0est}) is not 
better than $\sim$~20 -- 30\%\,. 

As was noted in Sec. 2.4, the redshift distribution 
of absorbers with larger $q_{thr}$ is weakly sensitive 
to small scale correlations of perturbations and 
it can be fitted by the expression (\ref{nw}). Indeed, 
for samples $Q_{14}^{20}$ and $Q_{14}^{50}$ these  
distributions are well fitted by expression (\ref{nw}) 
with $\langle q_{thr}\rangle\approx$ 0.035\,\&\,0.05, 
respectively, and the amplitude of the fit is only 
$\sim$~1.2 -- 1.3 times larger then what is expected in 
(\ref{nw}). However, already for the sample $Q_{14}^{10}$ 
the difference between expected and measured amplitudes 
of the fit (\ref{nw}) increases by up to $\sim$~1.75 times. 

These results show that the actual reliability and precision 
of our estimates (\ref{q0est}) are limited due to limited 
representativity of the samples and limited precision of 
estimates of the amplitude $\tau_0$ (\ref{tau0}). None the 
less, they suggest that: 
\begin{enumerate}
\item{} The redshift distribution of absorbers is quite 
well fitted by relations (\ref{nw}) and (\ref{nq0}) 
derived for the Gaussian initial perturbations.
\item{} The distribution of weaker absorbers 
is actually sensitive to the small scale cutoff  
of power spectrum described by the parameter $q_0\approx 
(0.9\pm 0.3)\cdot 10^{-2}$\,.
\item{} Comparison of estimates (\ref{est}) with 
$q_{thr}$ used for the sample selection indicates 
that, for weak absorbers with $q\sim 10^{-2}$, the 
expression (\ref{main}) with parameters (\ref{cosm}) 
underestimates the DM column density. However, for 
richer absorbers the divergence decreases to the 
quite moderate value ($\sim$~10\%). 
\end{enumerate}

The model as a whole 
and estimates of $q_0$ can be improved with richer sample 
of absorbers especially at high redshifts $z\geq$ 3.5 where 
the representativity of our samples is insufficient. However, 
as was noted in Songaila (1998) and Schaye et al. (2000), 
at $z\geq$ 3 the random variations of the UV background 
and the scatter of $\langle n_{abs}\rangle$ increase. 

\section{Summary and Discussion.}

In this paper we continue the analysis initiated in Paper 
I and Paper II based on the statistical description of 
Zel'dovich pancakes (DD99, DD02). This approach allows to 
connect the observed characteristics of absorbers with 
fundamental properties of the initial perturbations without any 
smoothing or filtering procedures, to reveal and to 
illustrate the main tendencies of structure evolution. It 
demonstrates also the generic origin of absorbers and 
the Large Scale Structure observed in the spatial galaxy 
distribution at small redshifts. 

We investigate the new more representative sample of 
$\sim$~4\,500 absorbers what allows us to improve the physical 
model of absorbers introduced in Paper I and Paper II and to 
obtain reasonable description of physical characteristics 
of absorbers. The progress achieved demonstrates again the 
key role of the representativity of observed samples for the 
construction of the physical model of absorbers and reveals 
a close connection between conclusions and observational 
database. 

However, the representativity of this sample is not 
sufficient for the more detailed study of structure evolution. 
Further progress can be achieved with richer sample of 
observed absorbers. 

\subsection{Main results}

Main results of our analysis can be summarized as follows:
\begin{enumerate}
\item The approach used in this paper allows to link the observed 
        and other physical 
        characteristics of Ly-$\alpha$ absorbers such as the 
        overdensity and entropy of the gaseous component and the 
        column density of DM component accumulated by absorbers.
\item The basic observed properties of absorbers are quite 
        successfully described by the statistical model of DM 
        confined structure elements (Zel'dovich pancakes). 
        Comparison of independent estimates of the DM 
        characteristics of pancakes confirms the self 
        consistency of the physical model for richer absorbers. 
        However, some characteristics of pancakes formation and 
        evolution remain uncertain.
\item   In the framework of this approach, all characteristics 
        of absorbers can be expressed through two functions, 
        $\xi\,\&\,\upsilon$, describing the systematic 
        and random variations of absorber properties. They are 
        directly expressed through the observed parameters, 
        $z,\,N_{HI}\,\&\,b$. 
\item  The main stages of structure evolution are 
        illustrated by separation of three subpopulations 
        of absorbers with high and low entropy and low 
        overdensity. 
\item The absorbers with high Doppler parameter, $b\geq$ 90 
        km/s, can be naturally identified with the embryos   
        of wall--like structure elements observed in the  
        spatial distribution of galaxies at small redshifts. 
\item The strong suppression of the mean number density of 
        absorbers at $z\leq$~1.8 can be naturally explained 
        by the influence of the observational threshold 
        $N_{HI}\geq~10^{12}{\rm cm}^{-2}$. 
\item Redshift distribution of weaker absorbers indicates 
        the probable systematic redshift variations of the 
        UV background. They can be caused by the 
        reionization of $HeII$.
\item The PDF of the DM column density and the redshift 
        distribution of absorbers are consistent with 
        Gaussian initial perturbations. 
\item We estimate the spectral moment $m_0$ which in turn  
        is linked with the cutoff of the power spectrum 
        caused by the mass of dominant fraction of DM 
        particles and the Jeans damping. 
\end{enumerate}

\subsection{Variations of intensity of the UV background}

The intensity of the UV background and its variations at 
redshifts $z\sim$~2 -- 3 were detected by different methods
(see, e.g., Songaila 1998; Scott et al. 2000; McDonald\,\&
\,Miralda-Escude 2001). However, the achieved precision 
of these estimates is limited. The variations of background 
temperature at these redshifts were also detected 
by Schaye et al. (2000) and McDonald et al. (2001).  

The redshift variations of the mean number density of 
weak absorbers discussed in Sec. 6.2 can be naturally 
related to such variations caused by the reionization of 
HeII and possible variations of activity of quasars and 
other sources of UV radiation. They can be enhanced by the 
incompleteness and limited representativity of our samples 
and by the nonhomogeneous redshift distribution of observed 
absorbers. 

However, these variations do not lead to appreciable  
variations of the mean absorbers characteristics discussed in 
Sec. 4 what indicates the complex character of absorbers 
evolution. The problem requires more detailed investigation 
using both observations and simulations. 

\subsection{Properties of absorbers} 

The physical model of absorbers introduced in Sec.2 links 
the measured $z$, $b$ and $lgN_{HI}$ with other physical 
characteristics of both gaseous and DM components forming 
the observed absorbers. Uncertainties in the available estimates 
of the background temperature and UV radiation and unknown 
parameters of the model restrict its applications. 
Fortunately, actions of these factors partly compensate 
each other, what allows us to obtain reasonable statistical 
description for majority of absorbers. The self consistency 
of this approach is confirmed by two independent estimates 
of the column density of DM pancakes. Numerical simulations 
can clarify action of unknown factors and improve the model. 

Analysis of mean absorbers characteristics performed in 
Sec. 4 shows that the sample of observed absorbers is 
composed of pancakes with various evolutionary histories. 
We discuss five main factors that determine absorbers evolution 
after formation. They are: the transverse expansion and 
compression of pancakes, their merging, the radiative 
heating and cooling of compressed gas, and the disruption 
of structure elements into a system of high density clouds. 
The first two factors change the overdensity of DM and gas but 
do not change the gas entropy. Next two factors change both 
the gas entropy and overdensity but do not change the DM 
characteristics. Using the entropy and overdensity we can 
roughly discriminate three subpopulations of absorbers,  
illustrate the influence of these factors and 
correlations between absorbers characteristics. 

Introduction of the {\it reduced} Doppler parameter, $\upsilon$, 
overdensity, $\Delta$, and entropy, $S$, allows to  
discriminate between the systematic and random variations 
of absorbers properties. The former ones are naturally 
related to the progressive growth with time of the DM 
column density of absorbers, $q(z)$, what can be described 
theoretically. On the other hand the action 
of random factors cannot be 
satisfactory described by any theoretical model. However, 
in the framework of our approach, the reduced parameters are 
found to be strongly correlated (\ref{ered}) and, in fact, 
the joint action of all random factors is summarized by 
one random function, $\upsilon$, directly expressed 
through the observed parameters (\ref{onbh}). These 
results alleviate the problem of absorbers description 
and, perhaps, the modeling of the Ly-$\alpha$ forest 
based on the simulated DM distribution (Viel et al. 2002). 
However, estimates of the matter fraction accumulated 
by absorbers (Sec. 4.4) demonstrate that the model 
needs to be improved. 

Estimates of the size of absorbers (Sec. 4.3) suggest a 
possible fast disruption of richer absorbers into a system 
of high density clouds, what is the first step of formation 
of dwarf galaxies. 

\subsection{Absorbers as elements of the Large Scale 
Structure of the Universe}

The close connection of absorbers with the Large Scale 
Structure (LSS) of the Universe observed in the spatial 
galaxy distribution was demonstrated in numerical 
simulations and was recently confirmed by direct 
observations (Penton, Stocke \& Shull 2002). Our results 
indicate the generic link of absorbers and DM Zel'dovich 
pancakes and demonstrate that the absorbers characteristics 
are consistent with Gaussian initial perturbations and 
the Harrison--Zel'dovich power spectrum. The possible 
identification of some absorbers with embryos of walls 
observed in the galaxy surveys at small redshifts shows 
that, perhaps, also a wider set of richer absorbers could 
be identified with the LSS elements at high redshifts. 

\subsection{Characteristics of the initial power spectrum}

The amplitude and the shape of large scale initial power 
spectrum are approximately established by investigations 
of relic radiation and the structure of the Universe at 
$z\ll$~1 detected in large redshift surveys such as the 
SDSS (Dodelson et al. 2002) and 2dF (Efstathiou et al. 2001). 
The shape of small scale initial power spectrum can be tested 
at high redshifts where it is not so strongly distorted by 
nonlinear evolution (see, e.g., Croft et al. 2002). 

Recent discussions on the mass of dominant fraction of DM 
component and the shape of small scale initial power 
spectrum are focused on the formation of low mass halos 
in CDM simulations in comparison with observed low mass 
satellites (see, e.g., Colin, Avila-Reese \& Valenzuela 
2001; Bode, Ostriker \& Turok 2001), and at the simulations 
of absorbers formation. Thus, Narayanan et al. (2000) 
estimate the low limit of this mass as $M_{DM}\geq$ 0.75 
keV, while Barkana, Haiman \& Ostriker (2001) increase 
this limit up to $M_{DM}\geq$~1 -- 1.25~keV. 

Here we can compare these estimates with  direct 
measurements of redshift distribution of absorbers. In fact, 
our estimates are related to two spectral moments of the 
initial power spectrum, $m_{-2}$ and $m_0$. The first of 
them depends mainly upon the large scale power spectrum 
and was independently estimated from the measured 
characteristics of observed galaxy distribution at small 
redshifts. Together with cosmological parameters, 
$\Omega_m$ and $h$, this moment defines the coherent length 
of initial velocity field, $l_v$, (\ref{mu}), used in Sec. 
2.3 for discussion of properties of DM distribution. 
The second moment, $m_0$, depends upon the small scale 
power spectrum and the shape of transfer function. This 
means that the estimates of the mass of DM particles are 
model dependent even though our estimates of $q_0$ and 
spectral moments are based on the analysis of observations.  

As was shown in DD99 and DD02, for the CDM and WDM models 
with the Harrison -- Zel'dovich asymptotic of power spectrum, 
$p(k)\propto k$, and transfer functions, $T(k)$, given in 
Bardeen et al. (1986) the coherent length of initial density 
perturbations, $l_\rho$, and the parameter $q_0$ can be 
written as follows:
\be
q_0 = 5{m_{-2}^2\over m_0},\quad l_\rho = q_0 l_v\approx 
0.34 h^{-1}{\rm Mpc}\left({q_0\over 0.01}{0.2\over\Omega_m 
h}\right)\,,
\label{q0}
\ee
\[
m_{-2} = \int_0^\infty dx~xT^2(x)\approx 0.023,\quad m_0 = 
\int_0^\infty dx~x^3T^2(x)\,,
\]
\[
x={k\over k_0},\quad k_0=\Omega_mh^2/{\rm Mpc},\quad 
l_v={1\over k_0\sqrt{m_{-2}}}={6.6\over \Omega_mh^2}
{\rm Mpc}\,, 
\]
where $k$ is the comoving wave number.
 
\begin{figure}
\centering
\epsfxsize=7.cm
\epsfbox{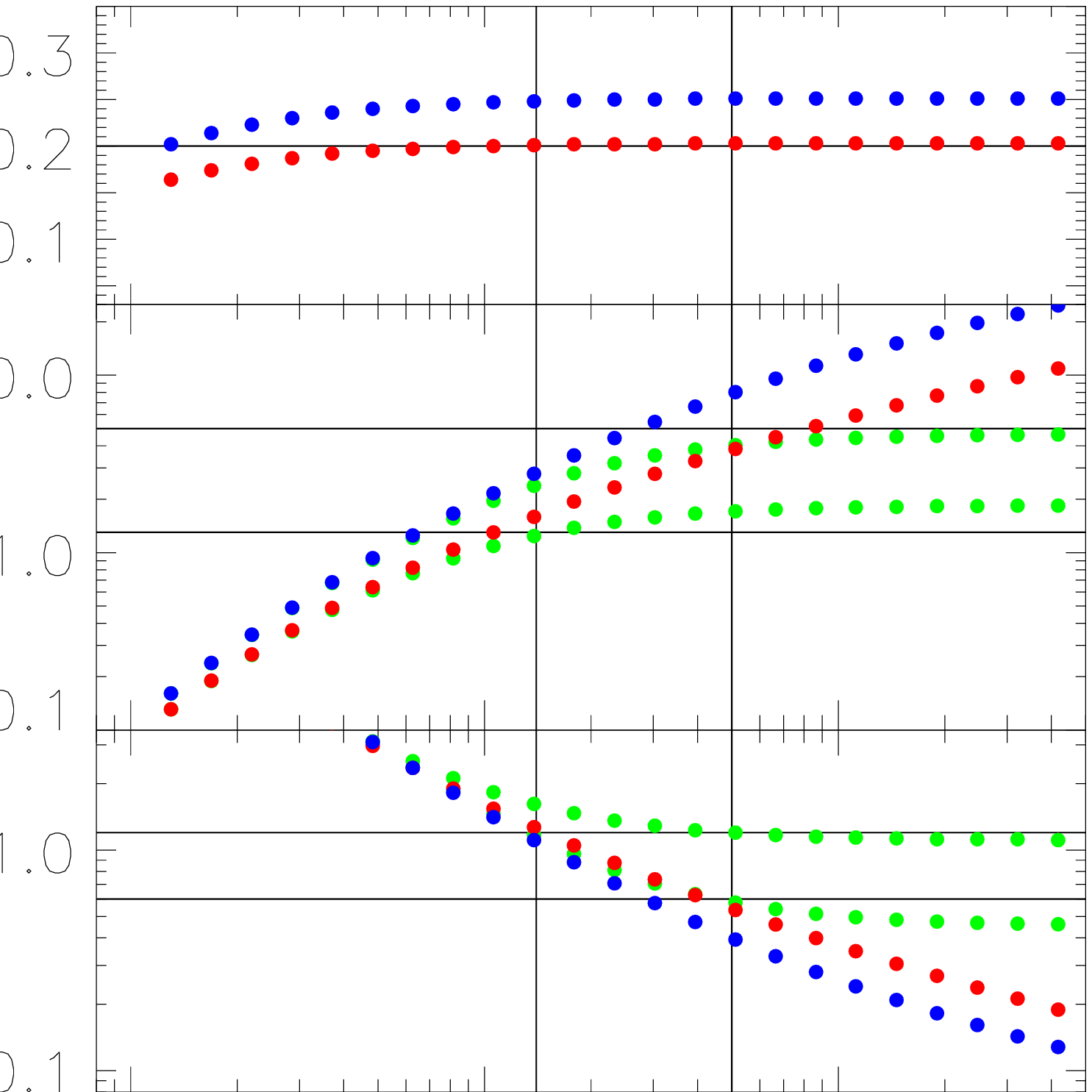}
\vspace{1.cm}
\caption{Spectral moments, $m_{-2}$ and $m_0$, and the 
parameter $q_0$ are plotted vs. $M_{DM}$ for the CDM (red points) 
and WDM (blue points) models with the damping factors $D_W$ 
(\ref{damp}) and with both damping factors, $D_W\,\&\,D_J$ 
(\ref{kkj}) for $x_J=(2\,\&\,0.5)\cdot 10^{-2}$, 
(green points). Black lines show the observational 
restrictions (\ref{qmm}).
} 
\end{figure}

For WDM particles the dimensionless damping scale, $R_f$, 
and the damping factor, $D_W$, are
\[
R_f=  0.2\left({\Omega_mh^2{\rm keV}}\over 
M_{DM}\right)^{4/3}\,,
\]
\be
D_W=\exp[-xR_f-(xR_f)^2]\,,
\label{damp}
\ee
(Bardeen et al. 1986). The Jeans wave number, $k_J$, and 
the damping factor, $D_J$, can be taken as 
\be
k_J^{-1}\approx 0.7 b_{bg}(1+z)H^{-1}(z),\quad 
D_J\approx (1+x_J^2x^2)^{-1}\,,
\label{kkj}
\ee
\[
x_J=k_0/k_J\approx 2\cdot 10^{-2}\sqrt{\Omega_mh^2
\over 0.13}\sqrt{4\over 1+z}\,{b_{bg}\over 16km/s}\,,
\]
(see, e.g., Matarrese\,\&\,Mohayaee 2002). 

Variations of the spectral moments and $q_0$ with the 
mass of DM particle obtained by integration of the 
power spectrum with these damping factors are plotted 
in Fig. 10 for the CDM and WDM transfer functions and 
for $x_J=2\cdot 10^{-2}\,\&\,0.5\cdot 10^{-2}$. As 
is seen from this Fig., the low limit $q_0\geq 0.07
\sqrt{x_J}$ caused by the Jeans damping (\ref{kkj}), 
restricts the range of masses of DM particles detectable 
with this approach to $M_{DM}\leq$~(3 -- 5)keV. 

Our 1$\sigma$ estimates 
\[
q_0\approx (0.6-1.2)\cdot 10^{-2},\quad m_0\approx 0.15-0.5,
\]
\be
M_{DM}\approx (1.5 - 5){\rm keV}\,,
\label{qmm}
\ee
are close to those of Narayanan et al. (2000) and Barkana, 
Haiman \& Ostriker (2001). They are based on absorbers with 
the redshifts $z\leq$~3.5 where the scatter of measured $n_{abs}$ 
is minimal and, so, they weakly depend upon both the Doppler 
parameter and velocities of absorbers which essentially 
distort $n_{abs}$ at higher redshifts (Sec. 4.4). 

There are plenty of possible candidates for the WDM particles 
with the mass $\sim$~1~keV. They are, for example, the sterile 
neutrinos, majorons and even shadow particles. Detailed review 
of possible candidates can be found in Sommer-Larsen \& Dolgov 
(2001) and Dolgov (2002). However, our estimates use the spectral 
moments and therefore do not forbid the existence of 
multicomponent heavy DM particles with both thermal and 
non-thermal (Lin et al. 2001) distributions with the same 
spectral moments $m_{-2}$ and $m_0$.

\subsection{Reheating of the Universe}

Recent observations of high redshift quasars with $z\geq$ 5 
(Djorgovski et al. 2001; Becker et al. 2001; Pentericci et 
al. 2001; Fan et al. 2001) demonstrate clear evidences in 
favor of the reionization of the Universe at redshifts 
$z\sim$~6 when the volume averaged fraction of neutral 
hydrogen is found to be $f_H\geq 10^{-3}$ and the 
photoionization rate $\Gamma_\gamma\sim (0.2 - 0.8)\cdot 
10^{-13}s^{-1}$ . These results are consistent with 
those expected 
at the end of the reionization epoch which probably takes 
place at $z\sim$~6. Extrapolation of the mean separation 
of absorbers discussed in Sec. 4.4 is consistent with these 
conclusions. 

\begin{figure}
\centering
\epsfxsize=7.cm
\epsfbox{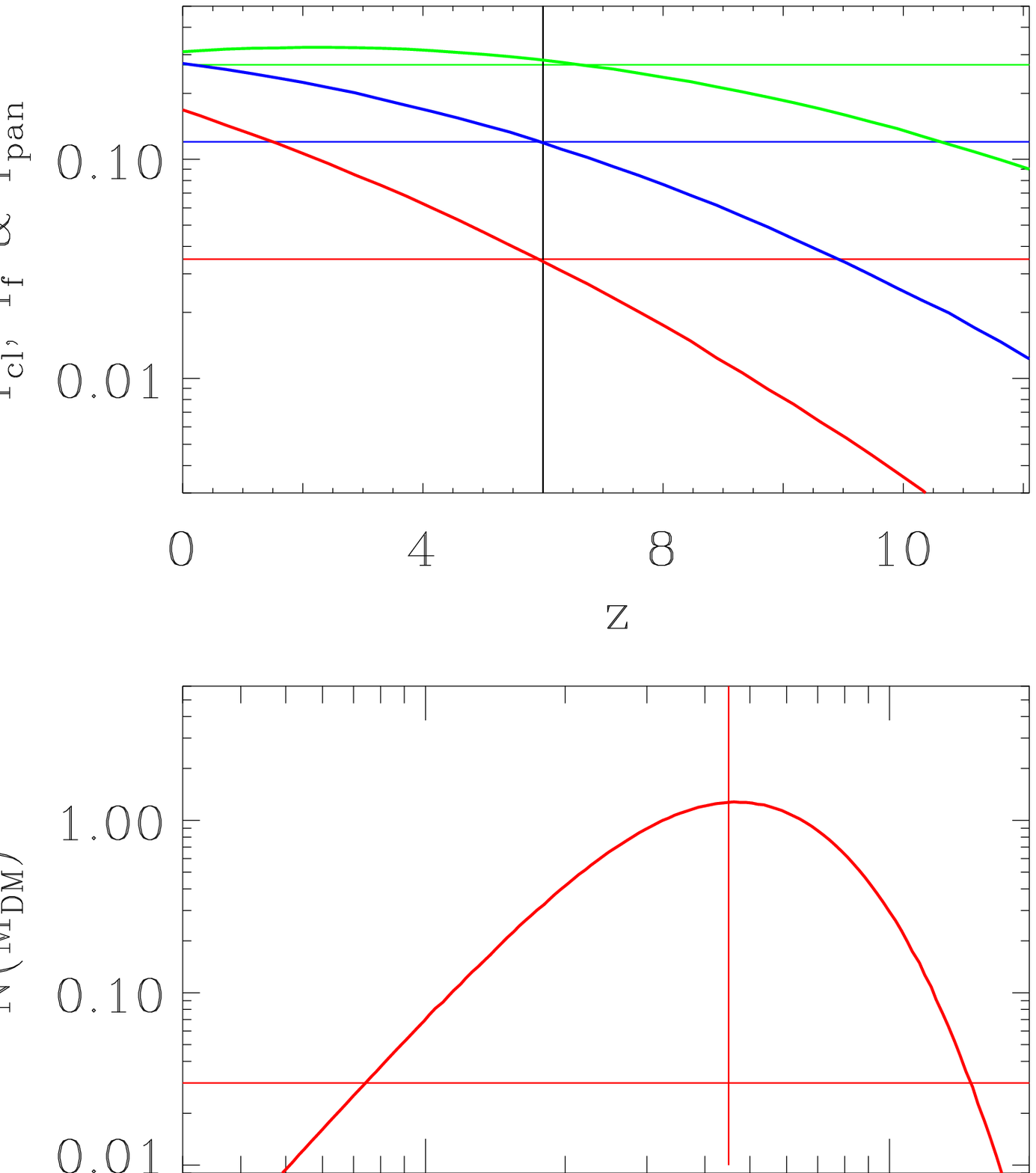}
\vspace{1.cm}
\caption{Top panel: expected redshift variations of the DM 
fraction accumulated by high density clouds, filaments and 
pancakes (red, green and blue lines), respectively, at 
$q_0=10^{-2}$. 
Bottom panel: the expected mass function of the DM clouds at 
$z=6$ and $q_0=10^{-2}$. 
} 
\end{figure}

These results can be compared with expectations of the 
Zel'dovich approximation (DD02) for $q_0=10^{-2}$. The 
potential of this approach is limited since it cannot 
describe the nonlinear stages of structure formation  
and, so, it cannot substitute the high resolution numerical 
simulations. However, it describes quite well many observed 
and simulated statistical characteristics of the structure 
such as the redshift distribution of absorbers and evolution 
of their DM column density. This approach does not depend on  
the box size, number of points and other limitations of 
numerical simulations and it successfully augments them.  

This approach allows to estimate the fractions of DM 
component accumulated by high density clouds, $f_{cl}$, 
filaments, $f_f$, and pancakes, $f_{pan}$, at different 
redshifts. These functions plotted in Fig. 11  for 
$q_0=10^{-2}$ show that at $z\sim$~6 only $\sim$~3.5\% 
of the matter is condensed within the high density 
clouds which can be associated with luminous objects. 
This value can increase up to $\sim$~5 -- 6\% with 
more correct description of the clouds collapse. At 
the same redshifts, $\sim$~27\% and $\sim$~12\% of 
the matter can be already accumulated by pancakes and 
filaments, respectively. These expectations are smaller  
than estimates of $\langle f_{abs}\rangle$ obtained in 
Sec. 4.4\,.

The Zel'dovich approximation allows also to estimate the 
mass function of all structure elements (DD02) at different 
redshifts. For $q_0=10^{-2}$ and at $z\sim$ 6, this function 
is plotted in Fig. 11. For these $q_0$ and $z$, the mean DM 
mass of structure elements is expected to be $\sim 10^{12}
M_\odot$ and the main mass is concentrated within clouds 
with $M_{cl}\sim 0.1\langle M_{cl}\rangle$. As is seen from 
Fig. 11, majority of the clouds have mass between $10^{-3}
\langle M_{cl}\rangle$ and 10 $\langle M_{cl}\rangle$. The 
formation of low mass clouds with $M_{cl}\leq 10^9 M_\odot$ 
is suppressed due to strong correlation of the initial density 
and velocity fields at scales $\leq l_\rho\sim 0.3h^{-1}$ Mpc 
(\ref{q0}). However, the numerous low mass satellites of large 
central galaxies can be formed in the course of disruption of 
massive collapsed clouds at the stage of their compression 
into thin pancake--like objects (Doroshkevich 1980; Vishniac 
1983). The minimal mass of such satellites was estimated 
in Barkana, Haiman \& Ostriker (2001).   

This means that the investigation of absorbers observed 
at high redshifts should be supplemented by the study of 
properties of dwarf {\it isolated} galaxies and discrimination 
between such galaxies and dwarf satellites of more massive 
galaxies. It seems to be a perspective way to discriminate 
between models with the Jeans damping (\ref{kkj}) and the 
damping caused by the mass of WDM particles, (\ref{damp}),  
and, in particular, between models with one and several 
types of DM particles.  

\subsection*{Acknowledgments}
AGD is grateful to Dr. S. Cristiani and Dr. T.S.Kim for the 
permission  to use the unpublished observational data.  
This paper was supported in part by Denmark's
Grundforskningsfond through its support for an establishment of
Theoretical Astrophysics Center and by the Polish 
State Committee for Scientific Research grant Nr. 2-P03D-014-17.  
AGD also wishes to acknowledge support
from the Center of Cosmo-Particle Physics, Moscow.
Furthermore, we wish to thank the anonymous referee for 
valuable discussion and many useful comments.

\end{document}